\documentclass[aps,prb,amsmath,amssymb,floatfix,12pt]{revtex4-1}
\usepackage{tabularx}
\usepackage{bm}
\usepackage{euscript}
\usepackage{graphicx}
\usepackage{color}
\usepackage{amsfonts}
\usepackage{exscale}
\usepackage{amsbsy}
\usepackage{subfigure}
\usepackage{textcomp}
\usepackage{comment}

\usepackage{hyperref}

\numberwithin{equation}{section}

\newcommand{\nn}{\nonumber}

\newcommand{\beq}{\begin{equation}}
\newcommand{\eeq}{\end{equation}}
\newcommand{\be}{\begin{eqnarray}}
\newcommand{\ee}{\end{eqnarray}}
\newcommand{\tr}{\textrm{tr}}
\newcommand{\CI}{{\cal I}}

\newcommand{\CM}{{\cal M}}
\newcommand{\CO}{{\cal O}}

\begin{document}

\title{ Enhanced Pairing of Quantum Critical Metals Near $d=3+1$}
\author{A. Liam Fitzpatrick$^{{\bar \psi},\psi}$, Shamit Kachru$^{{\bar \psi},\psi}$, Jared Kaplan$^{\phi_i^j}$, \\ S. Raghu$^{{\bar \psi},\psi}$, Gonzalo Torroba$^{{\rm tr}(\phi)}$, Huajia Wang$^{\bar{\psi}}$}
\affiliation{$^{\bar \psi}$Stanford Institute for Theoretical Physics, Stanford University, Stanford, California 94305, USA}
\affiliation{$^\psi$SLAC National Accelerator Laboratory, 2575 Sand Hill Road, Menlo Park, CA 94025, USA}
\affiliation{$^{\phi_i^j}$Department of Physics and Astronomy, Johns Hopkins University, Baltimore, MD 21218, USA}
\affiliation{$^{{\rm tr} (\phi)}$Centro At\'omico Bariloche and CONICET, Bariloche, Rio Negro R8402AGP, Argentina}

\date{\today}

\begin{abstract}

We study the dynamics of a quantum critical boson coupled to a Fermi surface in intermediate energy
regimes where the Landau damping of the boson can be parametrically controlled, either via large
Fermi velocity or by large $N$ techniques.  We develop a systematic approach to the BCS instability of such systems, including careful treatment of the enhanced log$^2$ and log$^3$
singularities which appear already at 1-loop.  These singularities arise due to the exchange of a critical boson in the Cooper channel and are absent in Fermi liquid theory.  We also treat possible instabilities to charge
density wave (CDW) formation, and compare the scales $\Lambda_{BCS}$ and $\Lambda_{CDW}$
of the onset of the instabilities in different parametric regimes.  We address the question of 
whether the dressing of the fermions into a non-Fermi liquid via interactions with the order
parameter field can happen at energies $> \Lambda_{BCS}, \Lambda_{CDW}$.

\end{abstract}

\maketitle

\section{Introduction}

Quantum critical fluctuations are believed to be responsible for the novel phenomena seen in many
modern materials, including the heavy fermions and the cuprates.  As one tunes to the quantum phase
transition, a critical order parameter field, taken to be  a scalar here,  can dress the Fermi liquid into a non-Fermi liquid, while at the same
time serving as additional pairing `glue' that can drive various instabilities, such as superconductivity
or CDW formation.  The interplay between these effects may yield a rich phase diagram; finding 
controlled approaches to map out such a phase diagram (starting with field-theoretic toy models)
may shed light on the enigmatic behavior of these materials. We focus on a class of quantum phase transitions in metals known as pomeranchuk instabilities, in which the bosonic order parameter fields condense at zero momentum; %the case where the most important boson modes have momentum low compared to $k_F$; 
our Lagrangian is
the one  relevant for Pomeranchuk instabilities (but not for transitions involving the onset of density wave order).  
%antiferromagnetic) phase transitions. 

The basic physics involves an interplay of several effects: the interaction with the Fermi surface can Landau damp the boson; boson exchange can dress the fermions into  --a non-Fermi liquid; 
and four-Fermi couplings controlling various instabilities can grow rapidly due to scalar exchange.  In order to find
tractable limits where we can parametrically determine which effects are dominant, we will introduce
several parameters: we will work in $d=3-\epsilon$ spatial dimensions, with a Fermi velocity $v_F$ and boson velocity $c$, and additionally introduce a parameter 
$N$.  In the theory with a given $N$, the boson is an $N \times N$ matrix, while we take the 
fermion to be in the fundamental representation; we note that this is quite different than the more commonly chosen vector-like large $N$ limit for the number of fermion flavors, and this matrix large $N$ limit can reveal different classes of possible behaviors.  For different choices of the global
symmetry group (SU(N) vs SO(N)),
we will see that different instabilities can dominate.  We will also see that $v_F / c$ seems to act as a control parameter, much like $N$, so that the theory simplifies at large $v_F / c$.  

In this paper, our philosophy will be the following.  We start at high energies with a Wilson-Fisher boson coupled weakly to a Fermi liquid.   We then follow the RG flow, decimating energy and momenta
(with a scaling we describe in more detail below) and lowering our ultraviolet cutoff as
$\Lambda = \Lambda_{UV} e^{-t}$.  We can associate definite scales $\Lambda_{BCS}$, $\Lambda_{CDW}$, $\Lambda_{NFL}$, and $\Lambda_{LD}$ with the phenomena of superconducting
pairing, CDW formation, emergence of the non-Fermi liquid, and Landau damping of the bosons.
We will choose regimes of $v_F/c$ and $N$ so that in each case, $\Lambda_{LD}$ falls beneath
a `dome' of either superconductivity or CDW formation.  A similar philosophy of studying 
quantum critical metals at intermediate
energies, perhaps governed by an approximate fixed point, has been followed in several earlier
works including \cite{SCS,GeorgesSachdev,Mahajan2013, Fitzpatrickone, FKKRtwo, Allais}.  

Previous authors already achieved rough estimates of some of these scales
in related problems \cite{Son1999,Shuster,Chubukov2005, Zaanen2008, Moon2010, Metlitski, Maier2014}.  
However, as has been appreciated for
some time \cite{Nayak1994a,Son1999,Schafer1999,Chubukov2005,Zaanen2008,Wang2013,Lee2013,Lee2009}, the presence of the scalar interacting with fermions at finite density leads
to novel features in the renormalization group, including the presence of ${\rm log}^2$
and ${\rm log}^3$ divergent diagrams that contribute to flows of
couplings (as opposed to the more normal logarithmic flow).  
We will generically refer to these as ``${\rm log}^2$ terms,'' for ease of expression.  Our main goal in this paper will
be to achieve a systematic (and hence extendable to higher orders) treatment of the renormalization group in the presence of these novel divergences.
To handle the ${\rm log}^2$ terms for the dominant four-Fermi couplings driving the BCS pairing and CDW formation, we will adopt the 
Wilsonian RG developed in [\onlinecite{WTF}]. A related RG approach to treat the tree level boson exchange has been proposed by Son\cite{Son1999}.

While we were thinking about these issues, interesting related works \cite{Metlitski, Gonzalo} appeared.  
These focus on parametric regimes distinct from the ones we consider, so while there is considerable
overlap of interests and philosophy, many detailed results are different.

\subsection*{The model}

We focus on the theory with Lagrangian $\mathcal L=\mathcal L_{\psi}+\mathcal L_{\phi}+\mathcal L_{\psi,\phi}$,
\begin{eqnarray}
\mathcal L_{\psi} &=&   \bar \psi^i \left[ \partial_{\tau}  + \mu_F -\epsilon_F(i  \nabla)  \right] \psi_{i} 
+ {\cal L_{\rm four-Fermi}}
\nonumber  \\
\mathcal L_{\phi} &=& {\rm tr}\left( m_{\phi}^2   \phi^2 + \left(\partial_{\tau} \phi \right)^2 + c^2 \left( \vec \nabla \phi \right)^2\right)   \nonumber \\
&&+  \frac{\lambda_\phi^{(1)}}{8 N} \tr (\phi^4) + \frac{\lambda_\phi^{(2)}}{8 N^2} (\tr (\phi^2))^2  \nonumber \\
\mathcal L_{\psi,\phi}&=& \frac{g}{\sqrt{N}}   \bar \psi^i \psi_{j} \phi^j_i\,.
\label{largeNaction}
\end{eqnarray}
 We consider fermions with a set of $N$ internal flavors 
$\psi_i, i = 1, \cdots N$, while the scalar $\phi^j_i$ is an 
$N \times N$ complex matrix.   Introducing the flavor degrees of freedom enables us to explore a very different asymptotic regime in this class of problems - where the bosonic degrees of freedom strongly overdamp the fermion modes - in a controllable fashion. This is the same theory as the one studied in ref. [\onlinecite{FKKRtwo}] and
parts of ref. [\onlinecite{Fitzpatrickone}], but here we will focus on subtleties of ${\cal L}_{\rm four-Fermi}$ that were not discussed in detail there.  This interaction depends on whether we consider $SU(N)$ or $SO(N)$ matrices $\phi$ (and hence whether $\psi$ and $\bar{\psi}$ are in conjugate representations, or the same real representations).  In the special unitary case, it is
\be\label{eq:4Fermidef}
{\cal L}^{SU}_{\rm four-Fermi} ~=~{u_{BCS }\over N} \bar\psi^i (k) \psi^i(p) \bar\psi^j (-k) \psi^j(-p) ~+ {{u_{CDW}} \over N} \bar\psi^i(k)
\psi^i(-k) \bar\psi^j(k) \psi^j(-k)~.
\ee
In this work we will focus on the corresponding two 4-Fermi channels that can lead to Fermi surface instabilities. 
It will be convenient to work with the dimensionless couplings
\be
\lambda_{BCS} \equiv u_{BCS} k_F^2 , \qquad \lambda_{CDW} \equiv u_{CDW} k_F^2.
\ee
 Due to the density of states, in our RG equations, $u_{BCS}, u_{CDW}$ will always enter in the above combinations.  This is part of an important general feature of our RG equations, that factors of $k_F$ never appear.  

In the special orthogonal or $SO(N)$ theory, the $\bar{\psi}$ and $\psi$ fields are both in the vector of $SO(N)$, and the product of two vectors contains a singlet.  So in addition to the i-j contractions visible in (\ref{eq:4Fermidef}), there is one more possible four-Fermi coupling:
\begin{equation}
\label{LfermiSO}
\Delta {\cal L}_{four-Fermi}^{SO} = \frac{u_{BCS2}}{N} \bar\psi^i(k) \bar\psi^i(-k) \psi^j(p) \psi^j(-p)
~.
\end{equation}
We will see later that there is a qualitative difference in the BCS running in these two theories.  

An important simplification is obtained by setting the first scalar quartic interaction
$\lambda_{\phi}^{(1)}$ to zero; this yields an enhanced symmetry only softly broken by the Yukawa
coupling  (i.e., the enhanced symmetry is broken only at $\mathcal O(1/N^2)$, and furthermore the Yukawa renormalizes four-$\phi$ scattering only through UV-convergent diagrams), and so is radiatively stable.  This makes the bosonic sector alone into a
vector-like $SO(N^2)$ Wilson-Fisher model, with a weakly coupled large $N$ fixed point.
Our UV fixed point consists of this and a decoupled Fermi liquid; we couple them through
the Yukawa $g$ which we will treat in
perturbation theory.

We adopt the same RG procedure as in ref. [\onlinecite{WTF}]; that is, at each step, we decimate boson modes with momentum $q$ in the range 
\begin{equation}
q \in (\Lambda_b - d \Lambda_b, \Lambda_b),
\end{equation}
as well as any remaining fermion modes with momentum $p$  in the range
\begin{equation}
p \in (\Lambda_f - d \Lambda_f, \Lambda_f),
\end{equation}
setting the independent cut-offs to satisfy $\Lambda_f/\Lambda_b$ fixed and small as they are lowered and we flow into the IR.   In this scheme, we are ensured that at any finite RG step, only high energy modes will be decimated.
Furthermore, logarithmic divergences that show up at tree-level in a partial wave basis for four-fermion interactions are absorbed into local counter terms, causing tree-level log running in this RG procedure.

\begin{figure}
\begin{center}
\includegraphics[width=0.50\textwidth]{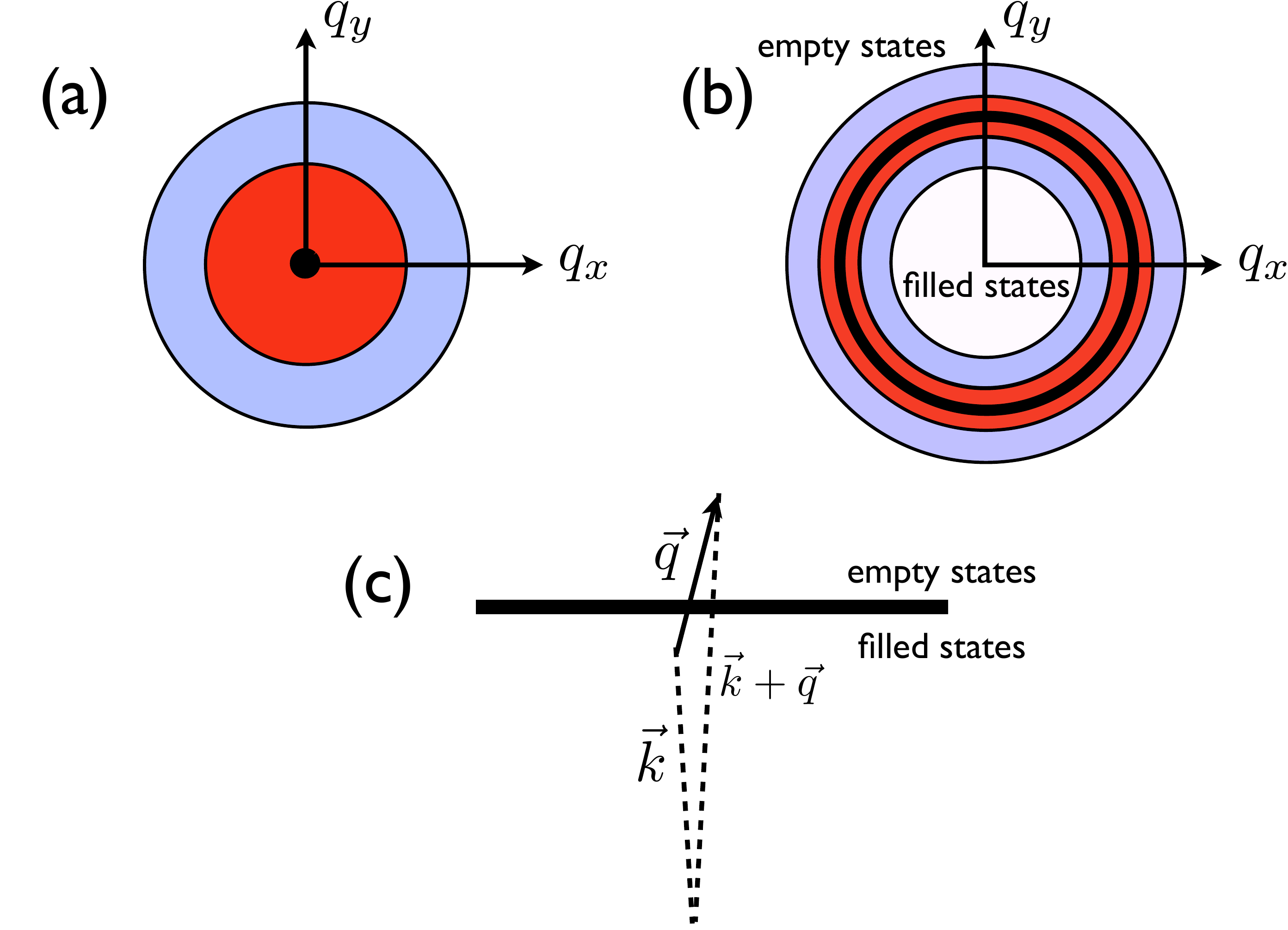}
\end{center}
\caption{Summary of tree-level scaling.  High energy modes (blue) are integrated out at tree level and remaining low energy modes (red) are rescaled so as to preserve the boson and fermion kinetic terms.  The boson modes (a) have the low energy locus at a point whereas the fermion modes (b) have their low energy locus on the Fermi surface.  The most relevant Yukawa coupling (c) connects particle-hole states  separated by small momenta near the Fermi surface; all other couplings are irrelevant under the scaling.   
}
\label{fig:scaling}
\end{figure}

We imagine tree-level scaling as in Figure \ref{fig:scaling}; this is what one would do to capture the decoupled Wilson-Fisher and
Fermi liquid fixed points\cite{Yamamoto2010}.  Note that in particular for fermions, only the component of momentum
perpendicular to the Fermi surface enters in the propagator and scales, while for the boson
the propagator is isotropic in momentum space.  This is appropriate for perturbation theory
around the UV decoupled fixed points.

The paper is structured as follows. First, \S \ref{sec:interp} discusses the BCS instability with gapless bosons at a heuristic level, highlighting the physical origin of multilogs and explaining physically how our RG will work.  The essential results of this paper are presented in this section with minimal technical details. The renormalization analysis is then carried out in \S \ref{sec:beta}, where we calculate the one loop beta functions for the BCS and CDW coupling. These results are applied in \S \ref{sec:phases} to the study of the phase diagram of the theory in terms of the different control parameters $N$, $v/c$ and $\epsilon$. Various more technical results are relegated to the Appendix. Furthermore, we describe in appendix \ref{app:alternate} how ${\rm log}^2$ terms can be treated in an RG procedure with solely loop-level log running, and discuss the disadvantages of such a procedure.

\section{Log squared divergences and their interpretation}\label{sec:interp}

\begin{figure}
\begin{center}
\includegraphics[width=1.0\textwidth]{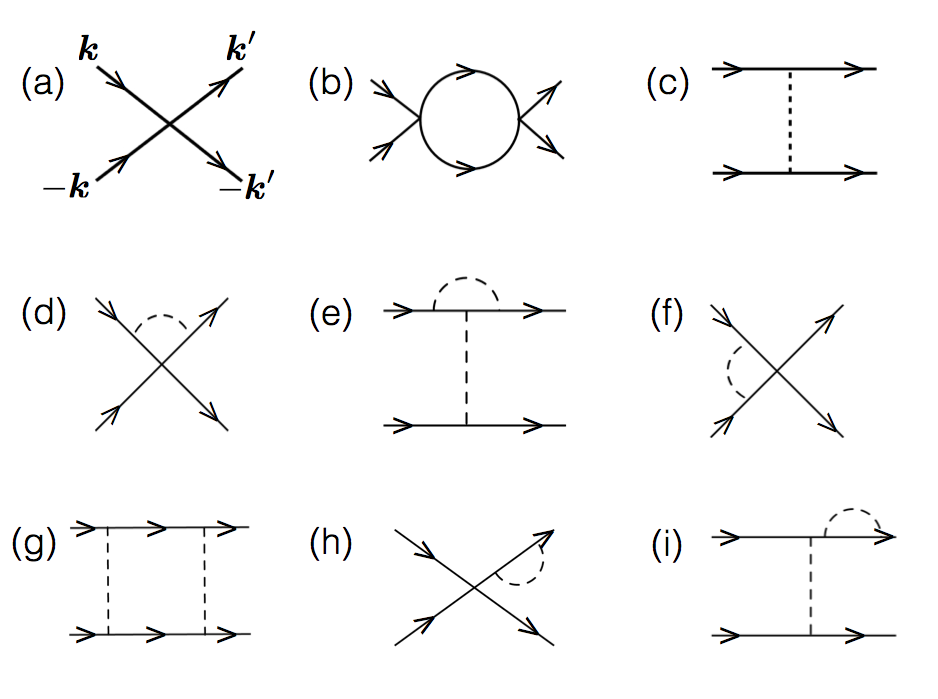}
\end{center}
\caption{Quantum corrections to the BCS 4-Fermi interaction.}
\label{fig:diags_sri}
\end{figure}

In this section, we consider the BCS instability of a Fermi surface coupled to a gapless boson at a heuristic level.  We first show how there are stronger divergences than those of ordinary BCS theory: in straight one-loop perturbation theory, the massless boson induces log-squared divergences in the Cooper channel (these are present in addition to the ordinary log divergences of Fermi liquid theory).  We then explain why such log-squared divergences pose a problem for the RG flows of the system, and how, if taken at face value, log-squared divergences invalidate notions of fixed points.  Lastly, we demonstrate how to interpret such divergences, using Wilsonian RG, decimating thin shells in momentum and energy space.

%\begin{figure}
%\begin{center}
%\includegraphics[width=0.6\textwidth]{ladders}
%\end{center}
%\caption{{\blue A naive estimate of the pairing scale is obtained from perturbation theory by summing up BCS ladder contributions to the renormalization of $\lambda_{BCS}$.  The second diagram is loq$^2$  divergent whereas the right most diagram is log$^4$ divergent.   } }
%\label{ladders}
%\end{figure}

The presence of log squared divergences in this problem is easily seen in direct perturbative calculations.  For simplicity, we consider here the effective interaction in the Cooper channel, with bare coupling constant $\lambda_{BCS}$.  When $g=0$, one recovers usual Fermi liquid behavior~\cite{Shankar, Polchinski}, and logarithmic divergences occur in perturbation theory from the diagram (b) in Fig \ref{fig:diags_sri}.  Resumming these diagrams amounts to doing one-loop RG and one recovers the usual scale where the BCS instability occurs.  

When the coupling to the scalar is non-zero, $g \ne 0$, however, new singularities are introduced in the Cooper channel.  There is a tree-level process, shown in Fig. \ref{fig:diags_sri}(c), which naively is non-singular, since it is simply a coupling function of the external momenta and frequencies.  The lowest order one-loop graphs involving the boson are in (d-g), which are discussed in more detail in appendix \ref{app:evaluation}.   Whereas (d) exhibits logarithmic divergences due to an effect very similar to a fermion-boson vertex correction, (e, f) exhibit log-squared divergences, and (g) exhibits a log-cubed divergence.  As mentioned in the text, we refer to both as ``log-squared" divergences, since as we shall see, they both have a common origin.

%Since both $g$ and $\lambda_{BCS}$ are classically marginal in $D=3+1$ in the decoupled fixed point, we first consider perturbation theory in both $g$ and $\lambda_{BCS}$, with $g^2 \sim \lambda_{BCS} \ll 1$.  %While both diagrams (d) and (f) are proportional to $g^2 \lambda_{BCS}$, (f) is more singular, and will cause breakdown of perturbation theory at a higher scale than (d).  
%{\blue In the standard approach to studying superconducting instabilities, one considers the renormalization of $\lambda_{BCS}$ from the particle-particle ladder diagrams, shown in Fig.\ref{ladders}.  Due to the critical nature of the boson here, however, the ladder summation has the form
%\begin{equation}
%\lambda_{BCS} \left[ 1 - c_1 g^2 \log^2{\left[ \Lambda/\mu \right]} + c_2 g^4 \log^4{\left[ \Lambda/\mu \right]}+ \cdots \right]
%\end{equation}
%where $\Lambda$ is an appropriate UV cutoff for the effective field theory, and $\mu$ is the energy scale of the external leg fermions, and $c_i$ are positive order unity constants that we ignore here in our heuristic discussion.  
%Thus, the naive expectation from this expansion would be that there is a parametrically enhanced breakdown scale $\mu_{BCS}$ due to the log-squared divergences:
%\begin{equation}
%\mu_{BCS} \sim \Lambda \exp{\left[ -1/\vert g \vert \right]}.
%\end{equation}
%}
This is distinct from the  naive expectations based on BCS theory, which would have predicted a substantially lower breakdown scale $\sim \exp{[-1/g^2]}$. 

Since both $g$ and $\lambda_{BCS}$ are classically marginal in $D=3+1$ in the decoupled fixed point, we first consider perturbation theory in both $g$ and $\lambda_{BCS}$, with $g^2 \sim \lambda_{BCS} \ll 1$.  While both diagrams (d) and (f) are proportional to $g^2 \lambda_{BCS}$, (f) is more singular, and will cause breakdown of perturbation theory at a higher scale than (d).  Thus, the naive expectation from perturbation theory would be that there is a parametrically enhanced breakdown scale $\mu_{BCS}$ due to the log-squared divergences:
\begin{equation}
\mu_{BCS} \sim \exp{\left[ -1/\vert g \vert \right]}.
\end{equation}
This is distinct from the  naive expectations based on BCS theory, which would have predicted a substantially lower breakdown scale $\sim \exp{[-1/g^2]}$.

The log-squared divergences, in addition to suggesting an interesting enhancement of superconductivity due to quantum critical fluctuations, pose a more fundamental challenge.  They cause the RG flows to depend explicitly (albeit still locally) on the energy scale.  This in turn has implications for universality  - RG flows with explicit dependence on energy scale (or equivalently, the RG time $t = \log{\left[ \mu/\mu(0) \right]}$) obtain a history dependence and generically weaken, if not completely destroy, the notion of fixed points where all history dependence must necessarily be lost.    The phenomena of log-squared divergences have occurred in several instances where fermions at finite density are coupled to gapless bosons.  A proper understanding of these divergences and their implications for universality remains incomplete.

These complications are resolved by considering the contribution from the tree-level process of boson-mediated scattering in the Cooper channel in Fig. \ref{fig:diags_sri} (c):

\begin{equation}
(c): V_{k,k'} = \frac{g^2 }{(k_0 - k_0')^2 + c^2 \vert \bm k - \bm k' \vert^2}\,,
\end{equation}
where $k$ and $k'$ are the external fermion momenta.
 It turns out that there is a logarithmic divergence hidden in this tree-level diagram, which contributes to the running of $V_{BCS}$.  The simplest way to see this is to consider decomposition into angular momentum harmonics:
\begin{equation}
V_L = \frac{1}{2} \int_{-1}^{1} d(\cos{\theta}) V_{k,k'} P_L(\cos{\theta}), 
\end{equation}
where $\theta$ is the angle between $\bm k$ and $\bm k'$.    This representation is appropriate for a rotationally invariant system;  representations of the crystalline point group should be used for a lattice system, but would not change the results below in a qualitative way.  

For simplicity, consider the $L= 0$ term; similar effects would occur for all $L$.  
The $L= 0$ amplitude is 
\begin{equation}
\label{V0}
V_0 = \frac{g^2}{2} \int \frac {d (\cos{\theta} )}{(k_0 - k_0')^2 + \vert \bm k - \bm k' \vert^2 }
\end{equation}
The idea \cite{WTF} (see also [\onlinecite{Son1999}, \onlinecite{Shuster}, \onlinecite{Metlitski}] for earlier developments towards this approach) is to treat the integral in Eq. \ref{V0} in a Wilsonian way, by decimating only fast bosonic modes.  To do this, define the momentum transfer $\bm q$ via
\begin{eqnarray}
q^2 &=& \vert \bm k - \bm k' \vert^2 \simeq 2k_F^2(1 - \cos{\theta}) \nonumber \\
d(q^2) &=& - 2k_F^2 d \cos{\theta}
\end{eqnarray}  
Thus, 
\begin{equation}
V_0 = -\frac{g^2}{4 k_F^2} \int \frac{ d (q^2) }{(k_0 - k_0')^2 + q^2} 
\end{equation}
In a Wilsonian treatment, the integration above is performed over a momentum shell $\left( \Lambda - d \Lambda \right) < q < \Lambda$.  We define $\delta V_0$ to be the contribution to $V_0$ from decimating this shell.  After making a change of variables $y = (k_0 - k_0')^2 +  q^2$, the integral becomes
\begin{eqnarray}
\delta V_0(c) &=& - \frac{g^2}{4  k_F^2} \int_{(k_0 - k_0')^2 + (\Lambda - d \Lambda)^2}^{(k_0 - k_0')^2 + \Lambda^2} \frac{ d y}{y} \simeq - \frac{g^2}{4 k_F^2} \frac{ d \left( \Lambda^2 \right)}{ \Lambda^2} \nonumber \\
&\simeq & - \frac{g^2}{2  k_F^2} \frac{ d \Lambda}{ \Lambda}
\end{eqnarray}
Since the boson exchange is singular at small angles, there is a log-divergence in (c) even though it is a  tree-level interaction.  This log-divergence is made manifest by working in the basis of angular momentum states.  It should be stressed that the presence of $k_F$ as a scale in the problem plays a crucial role in accommodating such a divergence.  It enables the conversion of an  angle-dependent scattering process in the BCS channel to a momentum transfer imparted to a scalar, which in turn can be subjected to Wilsonian mode elimination. 

 If a scale such as $k_F$ were absent, this log-divergence would not occur.  Take for instance the case of $\phi^3$ theory in the vicinity of the upper-critical dimension $5+1$ dimensions.  A tree-level diagram very similar to (c) occurs, which corresponds to  a $\phi^4$ interaction obtained by exchanging a ``fast'' mode.   However, since external modes correspond to ``slow modes", the fact that the locus of low energy states reside at $k=0$ implies that a ``fast" mode cannot couple slow modes.  For this reason, the dimension of the resulting $\phi^4$ term is identical to the one obtained by naive power counting.  

Having analyzed (c), we consider the contribution from (d) and (e)  to the running of $\lambda_{BCS}$.
 Let $\lambda_L$ now denote the spherical harmonic decomposition of the full BCS coupling, including the bare four-fermi coupling in the UV.
  In the Wilsonian treatment using the angular momentum representation, only (d) contributes:
\begin{equation}
d \lambda_L(d) = \lambda_L g^2 a_g \frac{d \Lambda}{\Lambda}
\end{equation}
where $a_g$ is a numerical constant.  To see why there is no contribution from (e), note that there is already a log-divergence present from the vertex correction.  Upon transforming to the angular momentum integral and decimating a momentum shell as above, one finds
\begin{equation}
d \lambda_L(e) =  \frac{g^4 }{4 k_F^2}a_g \left[ \frac{d \Lambda}{\Lambda} \right]^2
\end{equation}
 This does not contribute to the running to lowest order, because it is doubly small in the infinitesimal quantity $d \Lambda/ \Lambda$.  For similar reasons, (f) and (g) do not contribute at this order, since they are proportional to $(d \Lambda/ \Lambda)^2$ and $(d \Lambda/\Lambda)^3$, respectively.  
 
This illustrates the general treatment of multilogarithms: in our Wilsonian approach, multilogs are negligibly small and do not contribute, while their physics is encoded in the tree level running (c) in Figure \ref{fig:diags_sri}. This aspect of multilogarithms is also familiar from relativistic theories where, for instance, a $\log^2$ divergence in a two-loop diagram is cancelled by a one-loop counterterm. The novel point here is the appearance of running already at tree level.
 
In this simple Wilsonian theory, there is one last step: one must obtain the contribution from the rescaling of the fields.  In Fermi liquid theory, $\lambda_{BCS}$ is classically marginal.  However, this is no longer true when the fermions couple to the critical boson.  Near the upper-critical dimension,  the fermion self-energy exhibits logarithmic divergences,  which leads to an anomalous dimension for the fermion fields.  The anomalous dimension contributes to the running of $\lambda_{BCS}$:  this effect is given in diagrams (h) and (i).  
 
 Let us define the RG time $t = \log( \Lambda/\mu)$, where $\mu$ is the RG scale.  As the cutoff is lowered, shells are eliminated, and the change in $\lambda_L$ upon removing a thin shell of states is captured schematically by
 \begin{eqnarray}
 \lambda_L(t + \Delta t) &=& \left[ \lambda_L(t) +   \delta\lambda_L(c) + \delta \lambda_L (d) +  \delta\lambda_L (f)  \right] e^{- 4 \gamma_{\psi} \Delta t} \nonumber \\
    &=&  \left[  \lambda_L(t) - (a_1g^2 + a_2 \lambda_L^2 - a_3 g^2 \lambda_L) \Delta t \right] e^{- 4 \gamma_{\psi} \Delta t}
\end{eqnarray}
where the exponential factor above arises from the fermion anomalous dimension $\gamma_{\psi}$.  From this we read off the $\beta$-function for $\lambda_L$:
\begin{equation}
\label{beta_heuristic}
\beta_{ \lambda_L} = \frac{ d  \lambda_L}{d t} = -a_1 g^2+ (a_3 g^2- 4 \gamma_{\psi}) \lambda_L - a_2  \lambda_L^2 \,,
\end{equation}
 where $\gamma_{\psi} \sim g^2$, and $a_1, a_2, a_3$ are positive constants; more details are given in appendix \ref{app:evaluation}. Here we have neglected small $\mathcal O(g^4)$ corrections that will not play an important role in our analysis.
 
As the RG is carried out, and fast bosons are decimated, the effective interaction in the BCS channel continues to grow rapidly at small angles.  The procedure above is a systematic way to capture the singularity that results at small angles without integrating out gapless degrees of freedom.  The analysis here shows how the $\beta$-function obtained to one-loop order remains independent of the energy scale (or equivalently, the RG time $t$), and therefore notions of fixed points survive.  To go beyond this leading order, a more systematic field theoretic approach is the most natural way to proceed, which is developed in subsequent sections.  

At this point, the reader may wonder whether the simple RG analysis presented here leads to similar conclusions of an enhanced pairing scale, found from summing up ladder diagrams of Fig. \ref{ladders}.  We can address this issue by solving the flow equation for $\lambda_{BCS}$.  For purpose of illustration, we neglect the anomalous dimension contribution, and set $a_1 = a_2 = 1$.  Integrating both sides of Eq.\ref{beta_heuristic}, we find
\begin{equation}
\int_{\lambda_L(\Lambda)}^{\Lambda_L(\mu)} \frac{d \lambda}{-g^2 - \lambda_L^2} = \log{\left[ \Lambda/ \mu \right]}
\end{equation}
With the boundary condition $\lambda_L(\Lambda) = 0$, we find 
\begin{equation}
\tan^{-1}{\left[ \lambda_L(\mu)/ g \right] } = - g \log{\left[ \Lambda/ \mu \right]}
\end{equation}
A reasonable estimate for the pairing scale is obtained by requiring $\lambda_L(\mu_{BCS}) = - \infty$, leading to 
\begin{equation}
\mu_{BCS} = \Lambda e^{-\pi/2g }
\end{equation}
We see therefore, that even though the log squared divergent one-loop diagram did not contribute in to $\beta_{\lambda_L}$ its effects are completely captured by the contribution from the tree-level log divergence.  In this case, the conclusions from perturbation theory are completely consistent with the RG analysis.  
Here we have focused on the BCS four-Fermi coupling; we provide a similar discussion of the CDW coupling in appendix \ref{app:CDW}.

\section{$\beta$-function equations}
\label{sec:beta}

Let us now discuss the beta function for the BCS coupling, building from the previous analysis. 

\subsection{Renormalization}

It is useful to first explain in more detail the different RG scales that appear in the theory, and the general procedure that will be followed to calculate the beta functions. 

The low energy theory is controlled by two cutoffs~\cite{WTF}: $\Lambda_f$ determines the distance from the Fermi surface (its role is the same as the cutoff in the Fermi liquid RG of Shankar and Polchinski~\cite{Shankar, Polchinski}), whereas $\Lambda_b$ is the cutoff for the boson momenta. This cutoff also fixes the angular size $\Lambda_b/k_F$ of each patch on the Fermi surface, where interactions from boson exchange are important. The dominant quantum corrections are included if we choose $\Lambda_f , \Lambda_b < k_F$.
It is possible to study the bidimensional RG for couplings as $(\Lambda_f, \Lambda_b)$ are varied; however, for our purpose it will be enough to consider a one-dimensional version where both $\Lambda_f \sim \Lambda_b \propto e^{-t}$, the scale parameter of the RG transformation.

The presence of the additional cutoff $\Lambda_b$ that controls the decimation of high momentum bosons solves two problems that appear when the Fermi surface is coupled to a gapless boson: the presence of large logs proportional to $\log k_F$, and the appearance of multilogarithms.\footnote{The decimation of high momentum bosons through variations of $\Lambda_b$ is also important for canceling log divergences in nonlocal operators, thus ensuring the renormalizability of the theory [\onlinecite{WTF}].} In fact, these issues are deeply connected. To see this, let us analyze in more detail the diagram (f) in Fig. \ref{fig:diags_sri}.

If we take the limit where the external momenta are exactly on the Fermi surface, the loop integral takes the form
\be
\label{BCStriangle}
\CI &\equiv&  \int \frac{ d \omega d \ell  d^2 \hat{\Omega} }{(2\pi)^4} \frac{1}{(\omega^2 + \ell^2 v_F^2)(\omega^2 + \ell^2 + 2 k_F^2 (1-\cos \theta))}.
\ee
Although they do not appear here, the external momenta would serve as an IR regulator so that this integral is IR finite.  It is instructive to assume first that there is no cutoff $\Lambda_b$, so that the angular integration is over the whole Fermi surface. Performing this integral yields
\be
\CI &=& \int^{\Lambda_f}_\mu \frac{ d \omega d \ell }{(16 \pi^3)} \frac{1}{(\omega^2 + \ell^2 v_F^2)} \log \left(1 + \frac{4k_F^2}{\omega^2 + \ell^2} \right)
\label{eq:SampleLoopIntegral}
\ee
where we have introduced an IR regulator $\mu$, for simplicity.
The logarithm in the integrand is the `extra' log as compared to the BCS theory; it originates with the boson propagator integrated over the Fermi surface.

Performing the remaining integrals over $\ell$ and $\omega$ produces an additional logarithm, giving a leading dependence of $\log^2 \Lambda$ on the high-energy cutoff $\Lambda_f$.  In the limit $k_F \gg \Lambda$, we can easily evaluate the integrals, giving
\be
\CI = \frac{1}{16 \pi^2 v_F} \left( \log^2 \left( \frac{2k_F}{\mu}\right) - \log^2 \left( \frac{2 k_F}{\Lambda_f} \right) \right)= \frac{1}{16 \pi^2 v_F} \log \left(\frac{\Lambda_f }{\mu }\right)
   \log \left(\frac{4 k_F^2}{\Lambda_f  \mu }\right) .
   \label{eq:log2diagram}
\ee
In order for a renormalization scheme to work optimally, in general one wants to choose the RG scale in order to minimize the size of logarithms, since otherwise large logarithms lead to a breakdown of perturbation theory at higher orders.  However, in the above $\log^2$, there are two independent physical scales, the IR regulator $\mu$ (which should be thought of as the size of the external momenta) and the Fermi momentum $k_F$, and when they are very different this is no longer possible.  The culprit is the high-energy scale $k_F$ which appears in the above loop diagram due to exchange of high-momentum bosons between distant points on the Fermi surface.  This is resolved in terms of the additional cutoff $\Lambda_b$ that decimates high momentum bosons: $g^2 \log k_F$ is replaced by $g^2 \log \Lambda_b$, and this factor is interpreted as the tree level running of the 4-Fermion coupling.

We will study the RG of the system using renormalized perturbation theory, and working in the partial wave basis
\begin{equation}\label{eq:lambdaL}
\lambda_L = \frac{ 1}{2} \int_{-1}^{1} d(\cos{\theta}) \lambda_{BCS}(\theta) P_L(\cos{\theta}), 
\end{equation}
where $\theta$ is related to $q$ in the definition (\ref{eq:4Fermidef}) by $q=k_F \cos \theta$. For this, the `bare' coupling $\lambda_0$ that appears in the classical action is written in terms of the `renormalized' coupling $\lambda_L$ and counterterms,
\be\label{eq:lbare}
\lambda_{0,L}= \frac{ \lambda_L + \delta \lambda_L}{Z_\psi^2}\,.
\ee
Here $Z_\psi$ is the wavefunction renormalization for the fermions, whose RG variation gives the anomalous dimension $\frac{d}{dt} \log Z_\psi= -2 \gamma_\psi$. The counterterm $\delta \lambda_L$ cancels divergences from diagrams (b) -- (g) in Fig. \ref{fig:diags_sri}.  The cancellation of divergences is enforced at the RG scale $\mu \propto e^{-t}$, and the beta function then follows from (\ref{eq:lbare}) by noting that the bare coupling is independent of $t$. We next evaluate the quantum corrections to $\beta_{\lambda_L}$ explicitly.

\subsection{Calculation of the BCS $\beta$ function}

With the counterterm $\delta \lambda_L$ in place, we can already recognize how additional multilogs will arise: we will get the usual BCS log running from a fermion loop, but now at each vertex we can have a $\delta \lambda$ which by itself is already logarithmically divergent. This gives rise to double and triple logs, and we will show that they exactly cancel the corresponding logs from one loop diagrams with bosons and fermions. As a result, $\beta_\lambda$ will always be independent of $\log \Lambda$.

As we discussed in \S \ref{sec:interp}, the first divergence originates from the boson tree-level exchange in diagram (c); this requires a counterterm $\delta \lambda_L=g^2 \log (\Lambda/\mu)$, where we set $\Lambda_f \sim \Lambda_b \sim \Lambda$, and the tree level beta function is
\be
\beta_{\lambda_L}=\frac{d\lambda_L}{dt}=-g^2\,.
\ee
Recalling our convention that $\lambda_L<0$ is an attractive interaction, and that $t>0$ grows towards the IR, this implies that an attractive 4-Fermi coupling grows at low energies.

Next we analyze the one loop corrections. We will organize the calculation into two sets of diagrams in Fig. \ref{fig:diags_sri}: (b), (f) and (g) on the one hand, and (d), (e), (h) and (i) on the other.

First, diagram (g) gives a $\log^3$ divergence: we get $ \log \Lambda_f$ from the product of the two fermion lines, and a $(g^2 \log \Lambda_b)^2$ from the boson lines after transforming to the partial wave basis. Diagram (f) has a $g^2 \log\Lambda_b$ from the boson line, a $\log \Lambda_f$ factor from the integration over the two fermion propagators, and $\lambda_L + \delta \lambda_L$ from the vertex, so it gives rise to both $\log^2$ and $\log^3$ terms. Finally, diagram (b) has a $\log \Lambda_f$ from the fermion loop, times a $(\lambda_L + \delta \lambda_L)^2$ from the two 4-Fermi vertices. Combining these diagrams, all log-enhanced non-analytic functions of momenta are canceled, and the $\log^2$ and $\log^3$ contributions cancel out of the $\beta$ functions. 
As a result, the one loop contribution from (b)+(f)+(g) becomes
\be
\beta_{\lambda_L} \supset \,-\frac{1}{2\pi^2|v|}\,\frac{C_\lambda}{N} \lambda_L^2
\ee
where $C_\lambda$ is a group theoretic factor that depends on the channel.
Notice that this result agrees with the usual Fermi liquid running of the BCS interaction in the RG of~[\onlinecite{Shankar}, \onlinecite{Polchinski}]. What we have accomplished is to prove that the multilogs from exchange of light bosons cancel out in our RG equations.

Let us next consider diagrams (d), (e), (h) and (i). For clarity, we first discuss the theory with a singlet scalar, and afterwards include the group-theoretic factors in the matrix-valued scalar case.
The divergences from (h) and (i) are cancelled by the wavefunction factor $Z_\psi$ in (\ref{eq:lbare}), giving a contribution
\be\label{eq:anom1}
\beta_{\lambda_L} \supset -4\gamma_\psi \lambda_L\;,\;\gamma_\psi=\frac{g^2}{8\pi^2(1+|v|)}\,.
\ee
As expected, a positive anomalous dimension for the fermion tends to make the 4-Fermi interactions more irrelevant. Diagram (e) is not one-particle irreducible and its divergence is accounted for by the renormalization of the Yukawa interaction.  Diagram (d) contains a divergence $\sim \log \Lambda_b$ times a non-trivial function of momentum, which can be thought of as a sub-divergence coming from the boson exchange in the diagram times a loop integral.  
Its regularization and renormalization is taken into account by including our tree-level running of the four-fermi interaction, as is  carried out in detail in [\onlinecite{WTF}].
The result is that in the most convenient RG scheme, the vertex $\beta$ function takes the form
\be
\beta_{\lambda_L} &=& -a_1 g^2 -a_2 \lambda_L^2 \,,
\ee
where the anomalous dimension contribution $\sim g^2 \lambda_L$ exactly cancels the vertex contribution from diagram (d),  and $a_1, a_2$ are positive. There are various technical calculations that are involved in the cancellation of divergences and that are required to evaluate the coefficients $a_i$; these will however not be required for our following analysis, and so we refer the reader to [\onlinecite{WTF}] for more details.

In the matrix-valued scalar theory, the group theory factor for the anomalous dimension is the Casimir $C_2(\Box)$, and the vertex is proportional to $C_2(\Box)-\frac{1}{2} C_2(adj)$. (In particular, in our normalization for $SU(N)$ we have $C_2(\Box)=(N^2-1)/N$ and $C_2(adj)=2N$.) Now there is only a partial cancellation between the anomalous dimension and vertex corrections, leading to
\be
\beta_{\lambda_L} &=& -4 \gamma_\psi \lambda_L - a_1 g^2 - a_2 \lambda_L^2 + a_3 g^2 \lambda_L \,,
\ee
where all coefficients $a_i$ depend on $N$, and $a_1,a_2>0$; also $a_3$ is scheme-dependent.

For our analysis of the phase diagram below, the BCS beta function of the $SU(N)$ theory at large $N$ becomes
\be\label{eq:BCSlargeN1}
\beta_{\lambda_L}=-g^2-4 \gamma_\psi \lambda_L-\frac{1}{2\pi^2|v|}\,\frac{1}{N} \lambda_L^2\,.
\ee
For an attractive interaction $\lambda_L<0$, the first and third term lead to a rapid increase of $\lambda_L$, while the intermediate term tries to make $\lambda_L$ less relevant irrespective of its sign.  Crucially, the $\lambda_L^2$ term vanishes at $N\rightarrow \infty$, which implies that at large $N$ the BCS coupling remains perturbative up to scales $\sim e^{-\sqrt{N}/g}$ assuming $g^2 \lesssim 1/N$.  The $N$ in the exponent is an artifact of large $N$ in $SU(N)$, and we will see that for different global symmetry ($SO(N)$) the BCS instability reliably kicks in at scales $\sim e^{-1/g}$.  We discuss in detail the possibility of fixed points in section \ref{sec:RGE}. 

\subsection{CDW $\beta$ function}

The calculation for the CDW beta function is similar to the BCS case, but there are two important differences. First, the $1/N$ suppression factor 
for the $\lambda_L^2$ contribution to the beta function is absent, because the CDW condensate can be in a flavor-singlet channel. As a result, this term is not $1/N$ suppressed. We will further study the effect of this on the phase diagram in \S \ref{sec:phases}.

The other difference, as discussed in detail in appendix \ref{subsec:4Fermi1}, is that the CDW coupling only receives contributions from a small patch on the Fermi surface of angular size  
$ \sim \sqrt{2 \Lambda/ k_F}$ around the external fermion momentum. As a result, at a given RG time $t$, only the angular momentum modes with $t< 2 \log L$ will receive logarithmic contributions from diagrams such as (b) and (g) in Fig. \ref{fig:diags_sri}.
The beta function is then similar to (\ref{eq:BCSlargeN1}) after taking these points into account. In particular, at large $N$ it takes the form
\begin{eqnarray}
\label{FFrunsSU}
{d\lambda_{L, \rm CDW} \over dt} &\stackrel{t \lesssim 2 \log L}{\approx}& -g^2-4 \gamma_\psi \lambda_{L, \rm CDW}- \frac{1}{2\pi^2 |v|} \lambda_{L, \rm CDW}^2\,.
\end{eqnarray}

\subsection{Solutions to the RG Equations}
\label{sec:RGE}

An important qualitative aspect of the RG evolution above is that the four-fermion interactions tend to grow fairly quickly in the IR even though their anomalous dimension implies that they are irrelevant.  Thus in this case it is not sufficient to consider the scaling dimension of the interaction in order to understand if there is an instability. Such effects can in fact occur even in relativistic theories.  For example, a relativistic fermion and boson coupled through a Yukawa interaction $g \bar{\psi}\psi \phi$ has the following form of the one-loop $\beta$ function for the quartic $\lambda \phi^4$ coupling:
\be
\frac{d \lambda}{dt} &=& - \gamma_\lambda \lambda - a_\lambda \lambda^2 - a_g g^4,
\ee
where $a_\lambda, a_g>0$ are numeric constants,  $\gamma_\lambda = (4\gamma_\phi-\epsilon)$ is the renormalized scaling dimension of $\lambda$, and we recall that $t$ increases as we lower the cutoff. By a suitable choice of the space-time dimension and the couplings, one can take $\gamma_\lambda$ to be negative, so $\lambda$ is relevant and thus naively one would expect $\lambda$ to grow. But if $g^4$ is sufficiently large,  then the coupling $\lambda$ actually does the opposite and shrinks under RG flow. 
This is just the fact that when one is sufficiently far from a fixed point, the growth of interactions is not completely controlled by their scaling dimensions.  

What is notable about the RG of eqns. (\ref{eq:BCSlargeN1},\ref{FFrunsSU}) is that, since $\gamma_\psi \sim \CO(g^2)$,  the negative definite contribution $-(g^2 + \frac{1}{2\pi^2 v} \lambda^2)$ in the $\beta$ function tends to dominate over the anomalous dimension at all scales, for $d \approx 3$. 
Let us analyze this in more detail, focusing on the question of when the explicit scaling dimension for $\lambda$ can dominate the RG.  Consider the RG equation 
\be
\dot{\lambda} = -4 \gamma_\psi \lambda - g^2- \frac{a}{v} \lambda^2.
\label{eq:SimpleRG}
\ee
For generic initial conditions, the parameters $g,v,$ and $\gamma_\psi$ all run in this equation and a numeric solution is most straightforward. However, we can analytically see most of the content of this equation if we take $v\gg 1$ and rescale
\be
 g =  g' \sqrt{v}, \qquad \lambda' = \sqrt{a} \frac{\lambda}{g' v}, \qquad \textrm{ and } t' = \sqrt{a} t g'.
 \ee
  As described in the next section, $g'$ near $(2\pi) \sqrt{\epsilon}$ is then approximately constant as is $\dot{v}/v$, and the RG equation
can be seen to depend most essentially on the parameter
\be
c\equiv  \frac{2}{\sqrt{a}} \frac{1}{g'} \left( \gamma_\psi -  \frac{g'^2}{(4\pi)^2} \right).
\ee
Specifically, (\ref{eq:SimpleRG}) takes the form
\be
\dot{\lambda'} &=& -2 c \lambda' - 1-\lambda'^2.
\ee 

Now, consider for simplicity the case where the coupling $\lambda'$ starts out at $\lambda'(t'=0)=0$; for $c\ge 0$, even an arbitrarily large repulsive initial condition $\lambda' >0$ reaches $\lambda'=0$ in RG time of at most $t'\le \pi/2$ and so leads to essentially the same conclusions.  The solution to the RG equation is
\be
\lambda'(t') &=& -c-\sqrt{1-c^2} \tan\left( \sqrt{1-c^2} t' - \sin^{-1} (c) \right).
\ee
At $c=0$, this just gives $\lambda(t) = \frac{g' v}{\sqrt{a}}\lambda' =- \frac{g' v}{\sqrt{a}} \tan (\sqrt{a} g' t)$, which reproduces the result of [\onlinecite{Shuster}] after accounting for differing conventions. For $c<1$, this reaches a Landau pole at $t'=(1-c^2)^{-1/2}(\sin^{-1}(c) +\pi/2) = \frac{\pi}{2} + c + \CO(c^2)$.  On the other hand, when $c>1$, the coupling $\lambda'$ reaches a finite asymptotic value of $\lambda'(\infty) = -c+ \sqrt{c^2-1} = -\frac{1}{2c} + \CO(c^{-2})$.  Since $c \sim \gamma_\psi/g$, we see that for $\gamma_\psi/g$ small, the anomalous dimension loses and an instability develops, whereas for $\gamma_\psi/g$ large, the anomalous dimension eventually balances the other terms as $\lambda$ approaches a fixed point.  Furthermore,  the effect of the anomalous dimension starts to become visible over RG times of order $t' \sim 1/c$. Thus,  it is also the case that one requires $c \gtrsim 1$ for the fermions to get dressed into a Non-Fermi Liquid before the Landau pole occurs.  

Unfortunately, in the theories we consider, the anomalous dimension generated by the interaction with the boson is at most parametrically $\gamma_\psi \sim a g^2$, so that $c \sim \sqrt{a} g$.  We cannot make this ratio large within the controlled regime of perturbation theory in a simple and natural way.  However, it would be interesting to engineer additional perturbative contributions to $\gamma_\psi$ in order to break the relation between it and the size of the $g^2$ terms.   This could happen, for instance, in the vicinity of additional quantum critical points, or in the presence of emergent gauge fields.

\section{Control Parameters and Phase Diagram}\label{sec:phases}

In this section we discuss the parameters $\epsilon$, $v/c$, and $N$, and the group structure of the interactions. Then we study the phase diagram of our models in various controlled limits.

\subsection{ Large Fermi Velocity}

There will be two important new ingredients that we discuss carefully here, but that were not incorporated
into the previous work [\onlinecite{Fitzpatrickone}] and [\onlinecite{FKKRtwo}].
One is the possibility of a ${\it fast ~fermion}$ limit (also briefly discussed recently in [\onlinecite{Gonzalo}]), $v/c \rightarrow \infty$. For convenience, we will set $c=1$ as a choice of units. Because $g^2$ carries units of velocity$^3$, one must choose how to scale $g$ in the fast fermion limit.  To see how this should be done, consider that the fermion anomalous dimension  is
(for $N\ge 2$) 
\be 
2 \gamma_\psi = \frac{g^2}{(2\pi)^2 (|v|+1)}(1-N^{-2}).
\ee
If this is to have a finite limit, we must hold fixed the effective coupling
\be
\alpha &\equiv& \frac{g^2}{(2\pi)^2 (|v|+1)}.
\ee
Inspection of various other loop effects confirms that this is the natural ratio to hold fixed. Similarly, four-fermion interactions $\lambda$ should be scaled so that the ratio $\lambda /v$ is fixed, which can be read off from single boson exchange.

At $N=\infty$, the $\beta$ functions for $\alpha$ and $v$ in this large $v$ limit are
\be
\frac{d}{dt} \alpha = \alpha (\epsilon-\alpha), \qquad \frac{d}{dt} v = - \alpha v,
\ee
and thus $\alpha$ is driven toward the value $\epsilon$ under RG flow. The anomalous dimension at this point is $2 \gamma = \alpha = \epsilon$, which is {\it twice} the value of the anomalous dimension at the $N=\infty$ fixed point $g=g_*, v=v_*=0$.  This is possible because $v>0$ is not a fixed point, even though the $\beta$ function for $\alpha$ becomes arbitrarily small as $\alpha \rightarrow \epsilon$.

\begin{figure}
\begin{center}
\includegraphics[width=0.48\textwidth]{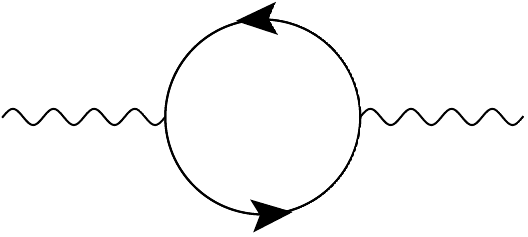}
\end{center}
\caption{The one-loop diagram generating Landau-damping of the boson.
}
\label{fig:LD}
\end{figure}

We will be concerned
with an intermediate energy regime where the Landau damping diagram of Figure \ref{fig:LD} is not an
important effect.  The standard result for Landau damping in $d=3$ is
\begin{equation}
\Pi(q_0, {\bf q}) ={1\over N} k_F^2 {g^2 \over 2\pi^2 v} \left[ {q_0 \over vq} {\rm tan}^{-1}\left({qv \over q_0}\right)\right]~.
\end{equation}

We notice that taking the limit $v \to \infty$ at fixed $x = q_0/q$,\footnote{Ref. \onlinecite{Gonzalo} confirmed that the $\beta$ functions and anomalous dimensions in the theory are dominated by regions of the relevant loop integrals when the boson $q_0$ is $\CO(q)$, rather than $\CO(v q)$, so holding $x$ fixed and $\CO(1)$ is the correct limit to take. } we find that
\begin{equation}
\label{hwoink}
\Pi(x) \sim  {1\over N} k_F^2 {\pi \alpha \over v} ~x~.
\end{equation}
We see that this intermediate energy range can be made parametrically large by taking large $N$,
as was exploited in earlier papers.  But we see
from (\ref{hwoink}) that it can also be made parametrically large at fixed $N$ by 
taking large $v$.
We will exploit both possibilities -- large $N$ and/or large $v$ -- when we consider different 
possible orderings of scales in our theory.

It is important to verify this large $v$ limit remains under control when we compute higher loop diagrams, or one-loop diagrams with more than two external bosons.  We verify the latter in appendix \ref{app:MultipleBosonScattering}, and argue that higher loop corrections should also be under control, although this remains to be confirmed in detail.

\subsection{Four-Fermi Terms}

Recall from the introduction that the four-Fermi Lagrangian in the special unitary case is
\begin{equation}
\label{LfermiSU}
{\cal L}_{\rm four-Fermi}^{SU} ~=~{\lambda_{BCS }\over k_F^2 N} \bar\psi^i (k) \psi^i(p) \bar\psi^j (-k) \psi^j(-p) ~+ {{\lambda_{CDW}} \over k_F^2 N} \bar\psi^i(k)
\psi^i(-k) \bar\psi^j(k) \psi^j(-k)~.
\end{equation}

The detailed difference between the two terms has important implications for the sizes of the 1-loop corrections to them.
The relevant one-loop diagrams appear in Figure \ref{fig:diags_sri}.  The first important point concerns the $N$-scaling
of various corrections. Consider diagram (b).  
We see that for the BCS interaction, (b) has no free indices on the internal loop -- it will give
corrections down by $1/N$. 
Therefore, the $\beta$-function
for BCS interactions in this theory at large $N$ is determined by tree-level running plus the wave-function renormalization
of the Fermi field. 

In contrast, for the CDW interaction, one should reverse one of the arrows on the fermion lines in
Figure \ref{fig:diags_sri}.  This makes a big difference in the $N$-counting in the special unitary theory.  
Now, diagram (b) gets a factor of $N$ from the loop, for instance.  This implies that
for small $g$ and ${\cal O}(1)$ values of the fermion couplings, the CDW instability will grow
much more quickly than the BCS coupling in the large $N$ SU(N) theory.  So there will be a region of
the phase diagram of this theory, as a function of the UV couplings, where the low-energy physics is governed by a charge density wave at sufficiently large $N$.

Turning to the special orthogonal case, the four-Fermi Lagrangian is 
\begin{equation}
\label{LfermiSO2}
\Delta {\cal L}_{four-Fermi}^{SO} = \frac{{\lambda}_{BCS2}}{k_F^2 N} \bar\psi^i(k) \bar\psi^i(-k) \psi^j(p) \psi^j(-p)
~.
\end{equation}
Here, there are singlets in the particle-particle channel, so diagram (b)
now contributes at leading order in $N$ to running of the BCS coupling $\lambda_{BCS2}$.  

The BCS beta functions were discussed in \S \ref{sec:beta}.  The CDW beta function differs in some interesting ways, as is discussed in detail in appendices \ref{app:CDW} and \ref{app:alternate}.
Here, we summarize the results.  The BCS and CDW beta functions for a large $N$ special unitary group are
\begin{eqnarray}
\label{FFrunsSU2}
{d\lambda_{L, \rm BCS} \over dt} &=& -4 \gamma\lambda_{L, \rm BCS}  - g^2 -  \frac{1}{2\pi^2 v} {1\over N} \lambda_{L, \rm BCS}^2 \\
{d\lambda_{L, \rm CDW} \over dt} &\stackrel{t \lesssim 2 \log L}{\approx}& -4 \gamma\lambda_{L, \rm CDW}- g^2 - \frac{1}{2\pi^2 v} \lambda_{L, \rm CDW}^2\
\end{eqnarray}

The relevant
equations for the $SO(N)$ case are
\begin{eqnarray}
\label{FFrunsSO}
{d\lambda_{L, \rm BCS2} \over dt} &=& -4 \gamma\lambda_{L, \rm BCS2} -g^2 - \frac{1}{2\pi^2 v}  \lambda_{L, \rm BCS2}^2 , \\
{d\lambda_{L, \rm CDW} \over dt} &\stackrel{t \lesssim 2 \log L}{\approx}& -4 \gamma\lambda_{L, \rm CDW} -g^2 -\frac{1}{2\pi^2 v} \lambda_{L, \rm CDW}^2 .\
\end{eqnarray}
Now both $\lambda_{BCS}$ and $\lambda_{CDW}$ have running due to self-couplings at leading order in $1/N$. 
In this case, the BCS instability always occurs first, because it has small $L$ modes that do not see any RG delay in their $\log^2$ terms, whereas the CDW couplings essentially exist in the effective theory only at large $L$.  This is in contrast to the $SU(N)$ case, where the $N$-suppression of the $\log^2$ terms for BCS can allow high-$L$ partial waves of CDW to compete with low-$L$ partial waves of BCS.  

An intriguing property of the SO(N) case is that, due to antisymmetry of the fermion wavefunction, the singlet channel (\ref{LfermiSO2}) requires an odd angular momentum $L$, while the flavor-antisymmetric piece --the first term in (\ref{LfermiSU})-- will be associated to even-$L$ modes. As a result, the large $N$ limit favors the superconducting instability of the $L=1$ mode over the more conventional $L=0$ channel. This provides a robust mechanism for p-wave superconductivity, and it would be interesting to analyze in more detail the low energy physics of the theory in this phase.

We stress that in order for our analysis to be valid over the full range of scales down to where the BCS or CDW instability sets in, one needs an $N$ or $v$ that is formally exponentially large in $1/g$, i.e. $N\gtrsim e^{{\rm const.}/g}$.  The reason is that Landau damping becomes large at a scale that is suppressed only by a power of $1/N$ or $1/v$, and our analysis applies only in the regime where it may be neglected.  In fact, once Landau damping becomes important, one would expect the CDW instability to turn off, based on the analysis in [\onlinecite{Shuster}].  The same is not true for the BCS instability, and a controlled analysis with a non-Fermi liquid near $3+1$ dimensions interacting with a Landau-damped boson and condensing into a superconducting state should be possible.\cite{treepee}

\subsection{Phase Diagrams}

Now that we have investigated both the RG behavior and the parametric scaling of our theory in the small couplings $\epsilon$, $1/N$, and $1/v$, we can describe the possible weakly coupled phase diagrams.

The basic parametric scalings we expect from the $\beta$-function equations is 
as follows.  In the $SU(N)$ theory, we expect
\begin{eqnarray}
\Lambda_{BCS}  &\sim& k_F ~e^{- {\sqrt{N v} \over g}}\\
\Lambda_{CDW} &\sim& k_F ~e^{-{\sqrt{v}\over g}}\\
\Lambda_{NFL} &\sim& k_F ~e^{-{v\over g^2}}\\
\Lambda_{LD} &\sim& {g k_F \over {N^{1/2} v}}~.
\end{eqnarray}
The parametric dependence on $N,g$, and $v$ is included above but not $\CO(1)$ constants that may appear in the exponent.
In the $SO(N)$ theory, we expect
\begin{eqnarray}
\Lambda_{BCS} &\sim& k_F ~e^{-{\sqrt{v} \over g}}\\
\Lambda_{CDW} &\sim& k_F ~e^{-{\sqrt{v} \over g}}\\
\Lambda_{NFL} &\sim& k_F ~e^{-{v\over g^2}}\\
\Lambda_{LD} &\sim& {g k_F \over {N^{1/2}v}}~.
\end{eqnarray}
As explained above, in the $SO(N)$ theory at large $N$ the superconductivity is dominated by the $L=1$ angular momentum channel, which may be interesting in connection to the phenomenology of unconventional superconductors.

\begin{figure}
\begin{center}
\includegraphics[width=0.65\textwidth]{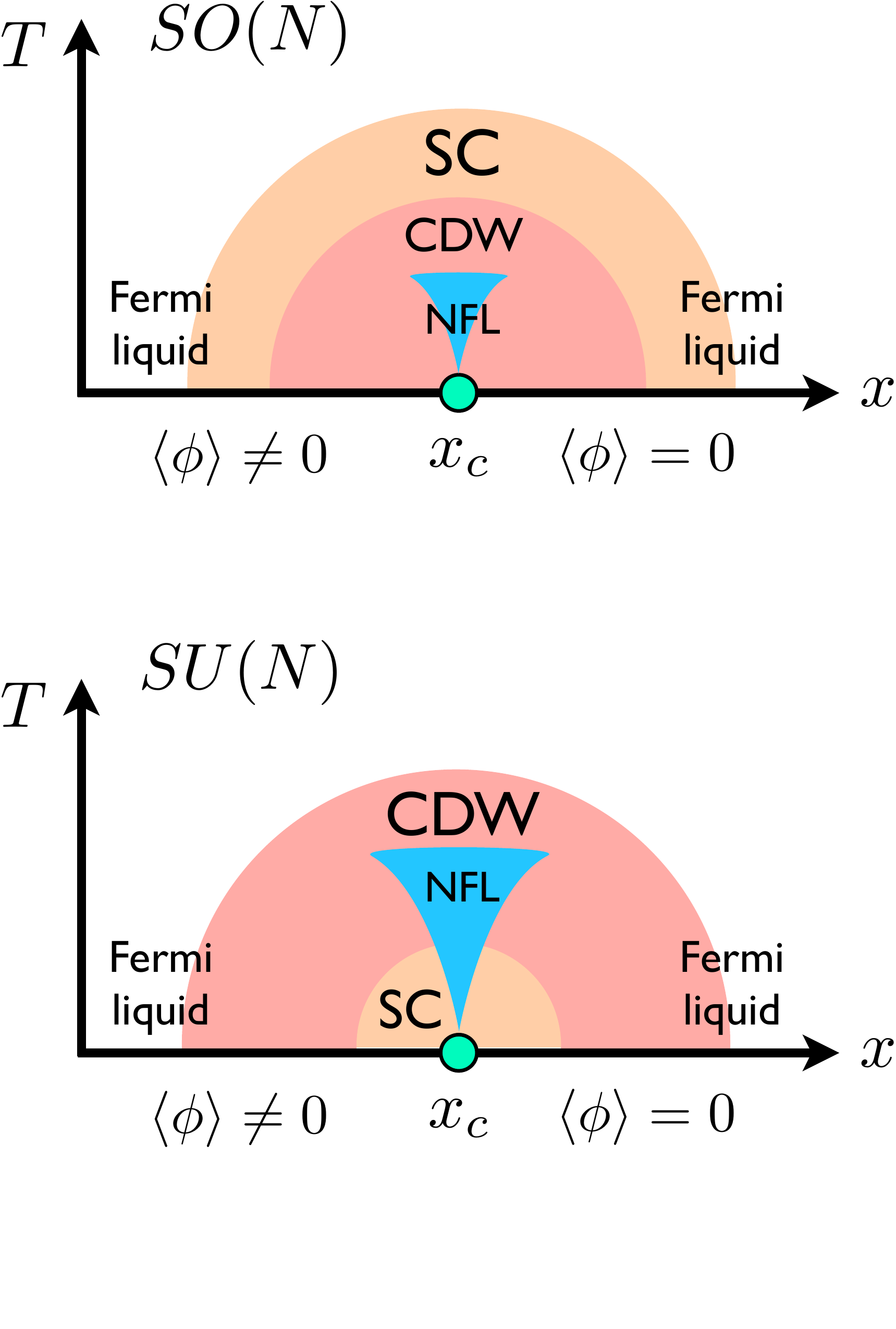}
\end{center}
\caption{An artist's portrayal of the competing phases of the $SO(N)$ ({\it top}) and $SU(N)$ ({\it bottom}) quantum critical metal. Essentially the same phase structure was found in [\onlinecite{Metlitski}], though in an analysis of a different theory. 
}
\label{fig:phases}
\end{figure}

This leads us to believe that the following structures are possible near $d=3$:

\noindent
$\bullet$ In the SU(N) theory, at exponentially large $N$ or $v$, one finds a flow from the UV fixed
point to a charge density wave, with the intermediate energy range governed by a $z=1$ boson coupled to a Fermi liquid.

\noindent
$\bullet$ In the SO(N) theory, at sufficiently large $N$ or $v$, one finds a flow from the UV fixed
point to a superconductor, with the intermediate energy range governed by a $z=1$ boson coupled to a Fermi liquid.
Possibly by varying parameters in a reasonable way, one can arrange a phase transition between
the superconductor and the charge density wave.

It is interesting to ask: can we find controlled theories which first dress the fermions into a non-Fermi liquid, and then condense into a superconducting phase?
It seems that by relaxing the constraint that we can ignore Landau damping of the boson (which restricted us to large $N$ or large $v$ in this paper), and studying the self-consistent
gap equations satisfied by both boson and fermion self-energies, we should be able to find consistent theories of non-Fermi liquids interacting with damped
bosons in a `Veneziano limit' (large number of fermion flavors $N_f$ and $N \times N$ matrix boson with fixed ratio $N_f/N$).    Quite plausibly there are limits in
the parameter space of these theories where the dressing into a non-Fermi liquid (and a damped boson) happens above a computable superconducting dome. This possibility was recently seen and briefly explored in [\onlinecite{Gonzalo}], and it would be clearly important to perform a more detailed analysis\cite{treepee}.

\acknowledgments{We thank  A. Chubukov, M. Mulligan, D. Son, and all the participants of the Stanford Non-Fermi Liquids Conference for interesting discussions about related
subjects.  S.K. is grateful to the Aspen Center for Physics for providing the rockies while he was
thinking about these issues.  SR is grateful to the Kavli Institute for Theoretical Physics, UC Santa Barbara for its hospitality during the time this work was carried out.  
This work was supported in part by the National Science Foundation grants PHY-0756174 (SK) and PHY-1316665 (JK), DOE Office of Basic Energy Sciences, contract DE-AC02-76SF00515 (SK and SR), a SLAC LDRD grant on `non-Fermi liquids' (ALF, SR, SK), the John Templeton Foundation (SK and SR), and the Alfred P. Sloan Foundation (JK and SR).  This material is based upon work supported in part by the National Science Foundation Grant No. 1066293 and PHY-1316665. ALF was partially supported by ERC grant BSMOXFORD no. 228169. GT is supported by CONICET, and PIP grant 11220110100752. }

\appendix

\section{Alternate RG Procedure}\label{app:alternate}

In the body of the paper, we have adopted the RG procedure developed in [\onlinecite{WTF}] using tree-level running.  To clarify the purpose of this procedure, it is helpful to contrast with a slightly more standard RG that involves only loop-level running, but then necessarily contains explicit dependence on RG time $t$ in the $\beta$ functions.  

To begin, let us return to the one loop result of diagram (c) of Figure 3, from equation (\ref{eq:log2diagram})
\be
\CI = \frac{1}{16 \pi^2 v_F} \left( \log^2 \left( \frac{2k_F}{\mu}\right) - \log^2 \left( \frac{2 k_F}{\Lambda} \right) \right)= \frac{1}{16 \pi^2 v_F} \log \left(\frac{\Lambda }{\mu }\right)
   \log \left(\frac{4 k_F^2}{\Lambda  \mu }\right) .
   \label{eq:log2diagram2}
\ee

This result has the crucial property that its derivative with respect to $\Lambda$ is completely independent of the low energy scale $\mu$.   In fact this was already manifest at the level of equation (\ref{eq:SampleLoopIntegral}), since the derivative of the integral would simply be the integrand evaluated at the cutoff, so that all low energy scales can be neglected.
The low energy scales will therefore disappear from the renormalization group equations.  Defining
for instance the bare and renormalized $\lambda_{BCS}$ in the standard way, by choosing the
counter-term to absorb the cutoff ($\Lambda$) dependent divergences, we see that
\begin{equation}
\beta_{\lambda} = {\partial \lambda \over \partial {\rm log} \Lambda}
\end{equation}
is ${\it independent}$ of $\mu$.  
This is not a coincidence, and is tied to a general property of the renormalization group --
that the couplings at a given choice of RG scale should not depend on external parameters such
as the momenta we choose for external particles.   So although we obtain an RG equation for the BCS and CDW couplings with explicit dependence on the logarithm of the cutoff, the crucial point that makes that RG sensible is that the derivative if $I_L$ with respect to the cutoff $\Lambda$ will only depend on $\Lambda$, the couplings, and the scale $k_F \gg \Lambda$, but it will not depend on any low-energy scales in the problem.
 
 We can now write down the beta function equations for the four-Fermi couplings
$\lambda_{BCS}$ and $\lambda_{CDW}$.  (We will not take care to get  $\CO(1)$ constants correct in this subsection.)
 Restricting to a spherical Fermi surface for simplicity, we again switch to spherical harmonics:
\be
\lambda(\theta) \Rightarrow \lambda_{L} \equiv \frac{1}{2} \int_{-1}^1 d \cos \theta \lambda(\theta) P_L(\cos \theta),
\ee
where $\theta$ is the scattering angle between the incoming particle pair and the outgoing particle pair. This also makes it much simpler to derive the form of the $\beta$ functions.  For this purpose, we can treat the boson exchanges in the triangle and box diagrams of 3b) and 3c)  as effective four-fermion interactions:
\be
V_{\rm eff}(\cos \theta) &=& g^2 \frac{1}{\omega^2 + (\ell^2+2k_F^2 (1-\cos \theta))} .
\ee
Then, the loop correction to $\lambda(\cos \theta)$ takes the partial wave form of 
\be
\delta \lambda_L &=& -4 \pi  \int \frac{d \omega d \ell}{(\omega^2 + v^2 \ell^2)} \left( \lambda_L^2 + 2 \lambda_L V_{L, \rm eff}+ V_{L, \rm eff}^2\right) ,
\ee
where $V_{L, \rm eff} = g^2\CM_L^{(g)}$ and is given at large $k_F$ by 
\be
V_{L, \rm eff} &=& \frac{g^2}{ 4} \left[ \log\left( \frac{ 4 k_F^2 }{ \ell^2 + \omega^2}\right) -2H_L \right],
\ee
where $H_L$ is the $L$-th harmonic number.
Changing coordinates to $\rho, \eta$ where $\omega = \rho \cos \eta,  \ell = \rho \sin \eta$, we can perform the $d \eta$ integration to obtain
\be
\delta \lambda_L &=&-\frac{8 \pi}{v}  \left[ \int^\Lambda \frac{d \rho}{\rho} \left( \lambda_L +  \frac{g^2}{k_F^2} (\log \frac{ 2 k_F}{\rho}- H_L)\right)^2 \right],
\ee
from which we can immediately read off
\be
\frac{d \lambda_L }{dt} &=& - \frac{8 \pi^2}{v}  \left( \lambda_L + \frac{g^2}{ k_F^2} t_L \right)^2 ,
\label{eq:PartialWaveRG}
\ee
where 
\be
t_L \equiv \log \frac{2k _F}{\Lambda} - H_L
\ee
is the RG time shifted by the ``delay'' $H_L$.

The advantage of this RG procedure over the one we adopt in the body of the paper is that it is somewhat more conventional, in that there is no tree-level running. However it is clearly somewhat unconventional in that there is explicit dependence on RG time in the $\beta$ functions.  More significantly, note that we could compute higher $k$-loop diagrams by combining $k+1$ factors of the vertex $\lambda_L + g^2 V_L$.  Since we have seen that each $V_L$ contributes parametrically as $g^2 \log (k_F / \Lambda)$, we will obtain an expansion in $g^2 \log \Lambda$.  This means that when this becomes order one, our naive perturbation theory will break down and we will need to resum to all orders in this quantity.  The RG procedure with tree-level running performs this resummation and thus has a greater regime of the theory under perturbative control.

\section{Evaluation of Diagrams (d-f) in Fig. \ref{fig:diags_sri}}
\label{app:evaluation}
We compute these diagrams with a UV cutoffs in momentum and frequency space $\Lambda$ ($c=1$).  
\subsection{Diagrams (d, e)}
This diagram has a logarithmic divergence arising from a process similar to the boson-fermion vertex correction.  Written explicitly, 
\begin{equation}
(d) = \lambda_{BCS} g^2 \int \frac{d^3 q dq_0}{(2 \pi)^4} D(q) G(k+q)G(k'+q)
\end{equation}
The log-divergence here comes in the limit of small momentum transfer in the Cooper channel: $\vert \bm k - \bm k' \vert \ll k_F$.  

Now, as found in [\onlinecite{Gonzalo}], the log divergence in the vertex correction comes multiplied by a singular function of $k-k'$. This divergence is cancelled by Diagram (e), which is similar to (d) with the replacement  $\lambda_{BCS} \rightarrow g^2 D(k-k')$. The renormalization of such nonlocal divergences is analyzed in detail in the companion work [\onlinecite{WTF}].

\subsection{Diagram (f)}

Since this diagram  produces a log-squared divergence, we evaluate it in explicit detail. 
\begin{eqnarray}
(f) = g^2 \lambda_{BCS} \int \frac{dq_0 d^3 q}{(2 \pi)^{4} } D(k-q)G(q)G(-q)
\end{eqnarray}
 The momenta in the boson propagator may be assumed to have magnitude $\vert \bm k \vert = \vert \bm q \vert \approx k_F$ (taking into account the differences $\vert \bm q \vert - k_F = \ell $ in the boson propagator lead to non-singular corrections):    
 \begin{eqnarray}
 D(k-q) \approx \frac{1}{(k_0-q_0)^2 + 2k_F^2(1-\cos{\theta})}
 \end{eqnarray}
 where $\theta$ is the cosine of the angle between $\vec k, \vec q$.  However, the dependence of the fermion propagators on $\ell$ produces the usual BCS logarithmic divergences and cannot be neglected.  and setting the measure $d^3 q = 2 \pi k_F^2 d \ell d \cos{\theta}$ (and defining $y = v \ell$),
\begin{equation}
 (f) = g^2 \lambda_{BCS} \frac{k_F^2}{v} \int \frac{d q_0 d y d \cos{\theta}}{(2 \pi)^3} \left[ \frac{1}{ (k_0-q_0)^2 + 2 k_F^2 - 2k_F^2 \cos{\theta}} \right] \left[ \frac{1}{(i q_0 - y)(-i q_0 - y)} \right]
\end{equation}
The integration over $\cos{\theta}$ involves only the first factor in brackets and produces a logarithm:
\begin{equation}
\int_{-1}^{1} \frac{d\cos{\theta}}{(k_0-q_0)^2 + 2k_F^2 - 2k_F^2 \cos{\theta}} = \frac{1}{2k_F^2} \log{\left[ \frac{(k_0-q_0)^2 + 4k_F^2}{(k_0 - q_0)^2}  \right]} \simeq \frac{1}{2k_F^2} \log{\left[ \frac{4k_F^2}{(k_0 - q_0)^2}  \right]} 
\end{equation} 
Thus, 
\begin{equation}
(f) = \frac{g^2}{2 v (2 \pi)^3} \lambda_{BCS} \int d q_0 d y \left[ \frac{1}{q_0^2 +y^2} \right] \log{\left[ \frac{4k_F^2}{(k_0 - q_0)^2}  \right]} 
\end{equation} 
The remaining integrals are naturally performed in  polar coordinates, defining $q_0 = r \cos{\phi}, y = r \sin{\phi}$ and the ``radius" $r$ has a UV cutoff $\Lambda$.  The integral becomes
\begin{equation}
(f) = \frac{g^2 \lambda_{BCS}}{2 v(2 \pi)^3}  \log{(\Lambda)} \int_0^{2 \pi} d \phi  \log{\left[ \frac{4 c^2 k_F^2}{\Lambda^2 \cos^2{\phi} } \right]} = -\frac{g^2 \lambda_{BCS}}{4 \pi^2 v}   \log{(\Lambda)} \log{\left[ \frac{ \Lambda }{2 k_F } \right]} + \cdots
\end{equation}
The last expression above show the explicit log-squared divergence.  

\subsection{ Diagram (g)}

It will also be useful to explain in some detail how the $\log^3$ dependence arises in the ``ladder'' diagram (g). For a Cooper-pair $\hat{n}(k_F+p',-k_F+p)$ scattering into $\hat{n}(k_F+p'+q,-k_F+p-q)$, the amplitude is given by
\be
\Gamma \sim -\frac{g^4}{(2\pi)^4} \int \frac{d\omega d\ell d^2 k}{(\omega^2+\ell^2+k^2)(\omega^2+(\ell+q)^2+k^2)\left(i\omega-v(\ell+p')\right)\left(i\omega+v(\ell-p)\right)}
\ee
We first evaluate the $\omega$ integral by residues. In order to see how the $\log^3$ behavior arises, it is sufficient to focus on one of the fermionic residues, say $\omega = -i v (\ell+p')$ for $\ell<-p'$. Multiplying this by two (which accounts approximately for the contribution from the other fermionic pole) and then integrating over $k$ obtains
\be
\Gamma
&\sim & \frac{g^4}{4\pi^2 v} \int ^{\Lambda_f}_{p'} \frac{d\ell }{2\ell - p'+p}\frac{1}{2\ell q - q^2}\log \frac{(1-v^2)\ell^2 - 2v^2 p' \ell -v^2p'^2}{(1-v^2)\ell^2+2\ell (q-v^2p')+q^2-v^2p'^2} + O(1/\Lambda_b)\nonumber\\
&\sim & F(p', p, q) + \mathcal O(1/\Lambda_f)\,.
\ee

The integral is UV convergent by power counting, and equals the finite answer $F(p',p,q)$. This is as the correction to the tree-level 4-Fermi vertex $\sim \frac{1}{q^2}$, and further log-divergences will be induced when we integrate over $q_\parallel$ to change to the spherical harmonic basis. For this we expand $F(p',p,q)$ for large q, and find
\be
F(p',p,q)\sim \frac{g^4}{4\pi^2 v}\frac{\log^2 q}{q^2} + \mathcal (\frac{\log q}{q^2}) \,.
\ee
Notice that we only introduced a component $q_\perp$ to make some of the expressions more tractable, but $q_\parallel$ will appear on the same footing in the $1/q^2$ factor, as this comes from the integral over the boson propagator. This result makes the $\log^3$ dependence manifest in the angular momentum basis.

\section{Pairing (BCS) and Forward Scattering (CDW)}\label{subsec:4Fermi1}
\label{app:CDW}

In the body of the paper, we mainly focused on the BCS four-fermi coupling; now we discuss the coupling relevant for CDW formation.
Note that because of the large energy cost of a boson mode which connects distant points on
the Fermi surface, the region which dominates the diagram comes from small values of $\theta$
(the angle between the initial and final points on the Fermi surface).  This is to be contrasted with the purely four-Fermi BCS one-loop diagram of Figure \ref{fig:diags_sri} (b).
There, the Feynman integral takes the form
\begin{equation}
\label{BCSfour}
\int ~{d\omega d{\ell} k_F^2 d^2\hat\Omega \over (2\pi)^4} {1 \over {(i\omega - v_F\ell) (-i\omega - v_F\ell)}}~.
\end{equation}
There is an integral which runs unconstrained over the whole volume of the Fermi surface $k_F^2 d^2\hat \Omega$.  This captures the physical point that in Fermi liquid theory, the BCS instability
can connect any pairs of antipodal points.

Now, let us consider the similar diagrams contributing to running of $\lambda_{CDW}$, which was defined below (\ref{eq:4Fermidef}). This channel is important due to the possibility of a charge density wave instability for the Fermi surface\cite{Peierls1930, DGR, Shuster}, namely the formation of a particle-hole condensate $\langle \psi^\dag(\vec k_F) \psi(-\vec k_F) \rangle$ of total momentum $2 k_F$. Quantum corrections to $\lambda_{CDW}$ are the same as the BCS diagrams in Fig. \ref{fig:diags_sri}, except that the arrow on one fermion line is reversed. Consider
the purely four-Fermi contribution from diagram (b) with the arrow on one line
reversed.  The integrand now takes the form
\begin{equation}
\label{DGRprocess}
\int ~{{d\omega} d^3{\bf p} \over (2\pi)^4} {1 \over {i\omega - (\epsilon_F( {\bf k_F} + {\bf p})- \mu_F)}} \cdot {1 \over {i\omega  - (\epsilon_F( {\bf k_F - p}) - \mu_F)}}~
\end{equation}
\begin{comment}
\begin{equation}
\label{DGRprocess}
\int ~{{d\omega} d^3{\bf p} \over (2\pi)^4} -{1 \over {i\omega - (\vert {\bf k_F} + {\bf p}\vert - k_F)}} \cdot {1 \over {-i\omega  + (\vert {\bf k_F - p}\vert - k_F)}}~
\end{equation}
\end{comment}
Here, we have been careful to indicate that the initial particle/hole pair is at momenta
$\pm {\bf k_F}$ on the Fermi surface - so there is a total momentum of $2 {\bf k_F}$ flowing
through the diagram.  The interesting point about (\ref{DGRprocess}) is that
the final states must also be close to $\pm {\bf k_F}$ to be states of the low-energy effective
theory.  In the integral over the loop momentum ${\bf p}$, we see that when we slide the
particle momentum by ${\bf p}$, we must shift the hole momentum also by ${\bf p}$.  This is
in contrast to the BCS process, where in the loop one shifts the two particles by ${\pm {\bf p}}$.  
As a result, while in the BCS process one can choose ${\bf p}$ vectors that transport one around
the Fermi surface and get contributions from the full Fermi surface, in (\ref{DGRprocess}) 
it is impossible to choose a macroscopic ${\bf p}$ keeping both a particle and a hole close to the Fermi surface.

This means that at an energy scale $\Lambda \ll E_F$, while the BCS diagram gets contributions from
the full area of the Fermi surface $\sim 4\pi k_F^2$, the CDW diagram is kinematically suppressed.  To see exactly what this suppression factor should be, let us take the incoming particle and hole momenta to be 
\be
\vec{p}_1 &=& \hat{\theta} k_F + \vec{q} \nn\\
 \vec{p}_2 &=& \hat{\theta} k_F - \vec{ q} 
 \ee
 where we have implemented the constraint that $\vec{p}_1+\vec{p}_2 = 2 {\bf k_F}$.  
 Now, a severe constraint comes from the fact that $q$ cannot take the fermions farther away from the Fermi surface than the cut-off $\Lambda$, i.e. $|p_1| < k_F + \Lambda$ and $|p_2| > k_F - \Lambda$:
 \be
k_F^2 + 2 q \cos \theta k_F + q^2 < (k_F + \Lambda)^2 \nn\\
k_F^2 - 2 q \cos \theta k_F + q^2 > (k_F - \Lambda)^2 
\ee
   At large $k_F$, this implies 
   \be
q \cos \theta \equiv \ell < \Lambda \textrm{ and }  q \sin \theta \equiv q_\parallel  < \sqrt{2 k_F \Lambda}
\label{eq:qparconstraint}
\ee
When we integrate over directions parallel to the Fermi surface, the constraint $q_\parallel \lesssim \sqrt{2k_F \Lambda}$ implies that the CDW loop corrections get a contribution from only a small piece of the full Fermi surface.  This is essentially a consequence of the same logic that implies that the only four-fermion interactions that are classically marginal are BCS scattering ($p,-p \rightarrow p', -p'$) and forward scattering ($p,p' \rightarrow p, p'$), since these are the only kinematics where all four momenta involved lie exactly on the Fermi surface.  Here, we are just being more precise about exactly how close to the Fermi surface the momenta must be for a given cut-off.  
For a fixed total momentum $\hat{\theta} k_F$, there is a single CDW scattering, from a particle-hole pair at momenta $(\hat{\theta} k_F, \hat{\theta} k_F)$ to itself, that lies exactly on the Fermi surface and thus is classically marginal all the way to zero energy.  However, at intermediate values of the cut-off, scattering from $(\hat{\theta} k_F, \hat{\theta} k_F)$ to $(\hat{\theta} k_F +\vec{q}, \hat{\theta} k_F - \vec{q})$ is close enough to the Fermi surface for sufficiently small $|q|$ that the corresponding coupling $\lambda_{\rm CDW}(\vec{q})$ behaves like a classically marginal interaction  until $\Lambda$ becomes too small, at which point it transitions to irrelevancy.

 This has important implications when we transform the CDW coupling to plane waves, 
\be
\tilde{\lambda}_{\rm CDW}(\vec{L}) &\equiv& \int d^2 q \lambda_{\rm CDW}(\vec{q}) e^{i \vec{L} \cdot \vec{q}/k_F} .
\label{eq:CDWpw2}
\ee
The reason is that the constraint $q_\parallel \lesssim \sqrt{k_F \Lambda}$ eliminates all modes with $|L| < \sqrt{k_F/\Lambda}$, retaining only modes with larger and larger $|L|$ as the cut-off is lowered.  Consequently, any fixed $\vec{L}$ CDW mode has a maximum amount of RG time available to it to develop an instability before it is essentially integrated out of the theory.

\subsection{Extension to CDW}
\label{sec:CDW}

Because of kinematic constraints, at low energies the CDW four-fermi interaction mediates only small-angle scattering.  For small variations around a fixed angle, the Fermi surface looks approximately flat and it is therefore simpler and more natural to  work with a basis of plane waves rather than spherical harmonics.  To be precise, let us denote the CDW four-fermi interaction for scattering from a particle-hole pair at momentum $(\vec{k}- \frac{\vec{q}}{2}, \vec{k}- \frac{\vec{q}}{2})$ to $(\vec{k}+ \frac{\vec{q}}{2},\vec{k}+ \frac{\vec{q}}{2})$ as $\lambda_{{\rm CDW}, \vec{k}}(\vec{q})$.  In the following, we will drop the dependence on $\vec{k}$, which can be thought of as the average momentum of a fixed patch.  We can approximately diagonalize the RG by taking a basis of ``plane waves'':
\be
\tilde{\lambda}_{\rm CDW}(\vec{z}) &\equiv& \int d^2 q \lambda_{\rm CDW}(\vec{q}) e^{i \vec{z} \cdot \vec{q}} .
\label{eq:CDWpw}
\ee

Consider first the RG in the case where $z^{-1} \ll \sqrt{k_F \Lambda}$, so that we can neglect the kinematic constraint on scattering angles.  The Fourier transform (\ref{eq:CDWpw}) turns the convolution from the one-loop diagrams into a product, so we can immediately write down the following simple form of the RG:
\be
\frac{d}{dt}  \tilde{\lambda}_{\rm CDW}(z) &=& \frac{2 \pi}{v}  \left( \tilde{\lambda}^2_{\rm CDW}(z) + 2 \tilde{\lambda}_{\rm CDW}(z) \tilde{V}_{\rm eff}(z,\Lambda) + \tilde{V}^2_{\rm eff}(z,\Lambda) \right)
\label{eq:CDWRGdiag}
\ee
where
\be
\tilde{V}_{\rm eff}(z,\Lambda) &=&g^2 \int d^2 k \frac{1}{\Lambda^2 + k^2} e^{i k \cdot z} = 2 \pi g^2 K_0(\Lambda z) 
 \ee
In this case, the double and triple logs arise because of the asymptotic behavior of the Bessel functions when $\Lambda \ll z^{-1}$
\be
K_0(\Lambda z) \stackrel{\Lambda z \ll 1}{\approx} - \log(\Lambda z/2) - \gamma_E
\ee
On the other hand, above the scale $z^{-1}$, the Bessel function dies exponentially, so there is effectively no contribution from the $\tilde{V}_{\rm eff}$ coupling:
\be
K_0(\Lambda z) \stackrel{\Lambda z \gg 1}{ \approx} e^{-\Lambda z} \sqrt{\frac{\pi}{2 \Lambda z}}
\ee

To consider the RG when $\Lambda \lesssim z^{-2}/k_F$ (i.e. $z^{-1} \gtrsim \sqrt{\Lambda k_F}$), we can go back to the integral $d^2 q$ and implement an upper bound through inserting a gaussian $e^{-q^2/(2\Lambda k_F)}$.  This produces for example
\be
\int  d^2 z d^2 y \tilde{\lambda}(z) \tilde{\lambda}(y) d^2 q e^{i q\cdot z - q^2/(2\Lambda k_F)} &=& \int d^2 z d^2y \tilde{\lambda}(z) \tilde{\lambda}(y) e^{- (y-z)^2 \Lambda k_F/2}
\label{eq:constraintRG}
\ee
 so the transformation $\lambda \rightarrow \tilde{\lambda}$ does a worse and worse job of diagonalizing the RG as $z^{-1}$ becomes greater than $\sqrt{\Lambda k_F}$.  Instead, the kinematic constraint acts like a ``box'' that contains modes only with  $z \gtrsim (\Lambda k_F)^{-1/2}$, as can be seen explicitly by choosing a basis that diagonalizes the LHS of (\ref{eq:constraintRG}).  In physical terms, the cut-off $ \sqrt{\Lambda k_F}$ on parallel scattering momentum integrates out modes below this ``$z$"-wavelength.''

Let us summarize the RG strategy for $\lambda_{\rm CDW}$ in this basis.  We choose a value of $z$ that is sufficiently large that it is bigger than $1/\sqrt{\Lambda k_F}$ over the entire range of the RG that we are going to use.  Then, the RG equations are diagonal to very good approximation, and we can use (\ref{eq:CDWRGdiag}).  As we run down in scale, the contributions from the Yukawa through $\tilde{V}_{\rm eff}(z)$ start out exponentially suppressed and are thus zero for all practical purposes.  Once we flow $\Lambda$ down to $z^{-1}$, the exponential suppression turns off and we see the double and triple logs, which cause a quick growth in $\tilde{\lambda}$.  We can keep flowing $\Lambda$ down to the scale $z^{-2}/k_F$ and look for a strong coupling scale and the onset of a CDW instability; however, below this scale there are no longer any modes that can scatter through $\tilde{\lambda}_{\rm CDW}(\vec{z})$, and it effectively gets integrated out of the theory.

\section{Boson Scattering via Fermion Loops and large $v_F$ limit}
\label{app:MultipleBosonScattering}

We would like to study the scattering of scalar bosons with speed of sound $c=1$ via fermion loops, in order to understand its parametrics in $g$ and $v_F$.  The 4-boson diagram is
\be
&& g^4 k_F^2 \int \frac{d \omega d\ell d \Omega}{(2 \pi)^4}
\left( \frac{1}{ i \omega - v_F \ell }  \right)
\left( \frac{1}{ i (\omega - E_1) -  v_F ( \ell - \hat \Omega \cdot k_1) }  \right)
\nn \\
&& \times
\left( \frac{1}{ i (\omega - E_1 - E_2) - v_F ( \ell - \hat \Omega \cdot (k_1 + k_2))}  \right)
\left( \frac{1}{ i (\omega +E_3) -  v_F (  \ell +  \hat \Omega \cdot k_3)} \right)
\nn
\ee
In order for the integral to be non-vanishing we must have poles on both sides of the contour in $\omega$ and $\ell$.

We can choose the kinematics so that
\be
k_{1/2} = (E, \pm E, 0, 0)
\ee
for simplicity, so that we can write the integral as
 \be
&& g^4 k_F^2 \int d \omega d\ell d \Omega
\left( \frac{1}{ i \omega - v_F \ell }  \right)
\left( \frac{1}{ i (\omega - E) -  v_F ( \ell - \hat \Omega \cdot k_1) }  \right)
\nn \\
&& \times
\left( \frac{1}{ i (\omega - 2 E) - v_F  \ell }  \right)
\left( \frac{1}{ i (\omega +E) -  v_F (  \ell +  \hat \Omega \cdot k_4)} \right)
\nn
\ee
If we close the contour in the upper half $\ell$ plane we get a contribution from each of the four poles.
We know that the sum of all four terms must vanish for any given $\omega$ where they all contribute, because it is the contour integral about all the poles (in $\ell$)  simultaneously, and so it can be deformed to enclose none of the poles.  Thus we can write the result of the $\ell$ integral as an integral in $\omega$ over a finite interval.  These $\omega$ integrals are then trivial, and so we find
\be
&& \frac{g^4 k_F^2}{v_F} \int d \Omega 
\left[
2 E \left( \frac{1}{ -i E +  v_F ( \hat \Omega \cdot k_1) }  \right)
\left( \frac{1}{ - 2 i E}  \right)
\left( \frac{1}{ i (E) -  v_F ( \hat \Omega \cdot k_4)} \right) \right.
\nn \\ && 
+ E \left( \frac{1}{ E^2 + v_F^2 (\hat \Omega \cdot k_1)^2 }  \right)
\left( \frac{1}{ 2 i E -  v_F (\hat \Omega \cdot k_1 + \hat \Omega \cdot k_4)} \right)
\nn \\ && 
+ \left. 3 E \left( \frac{1}{ i E  - v_F \hat \Omega \cdot k_4 }  \right)
\left( \frac{1}{ 3 i E -  v_F  \hat \Omega \cdot k_4 }  \right)
\left( \frac{1}{ 2 i E - v_F  (\hat \Omega \cdot k_1 + \hat \Omega \cdot k_4) }  \right) \right]
\ee 
Each of the terms appears to be suppressed as 
\be
\frac{g^4}{v_F^3}
\ee 
in the limit of large $v_F$, but this is a bit too naive.  Note that for special $\hat \Omega$ and $k_i$ we can get an integrand of order $g^4/v_F$ by having $\hat \Omega \cdot \vec k_1 = \hat \Omega \cdot \vec k_4 = 0$, so we need to be careful.
 
Studying the first term as an example, we can write it as
\be
\frac{g^4 k_F^2}{v_F} \int d \cos \theta d \phi
 \left( \frac{i}{ -i E +  v_F ( | \vec  k_1| \cos \theta) }  \right)
\left( \frac{1}{ i E -  v_F ( k_{4x} \cos \theta + k_{4y} \sin \theta \cos \phi)} \right) 
\ee
The scenario with the fewest factors of $v_F$ would be one where $k_1$ and $-k_4$ are aligned, so that $|\vec k_1 | = | \vec k_4 | + \mathcal{O}(\frac{1}{v_F})$ and we obtain an integral
\be
&& \frac{g^4 k_F^2}{v_F} \int d \cos \theta d \phi
 \left( \frac{i}{ -i E +  v_F ( | \vec k_1| \cos \theta) }  \right)
\left( \frac{1}{ i E +  v_F ( | \vec k_1| \cos \theta)} \right) 
\nn \\
&=& \frac{g^4 k_F^2}{v_F} \int_{-1}^1 d x
 \left( \frac{i}{  E^2 +  v_F^2 | \vec k_1|^2 x^2 }  \right)
 \nn \\
&=& \frac{g^4 k_F^2}{ v_F^2}   \frac{2 \tan^{-1} \left( \frac{v_F |k_1| }{ E} \right)}{E |\vec k_1|^2}
\ee
This result is of order $g^4 / v_F^2$, but it can only be obtained in a region of phase space where the angles of the $\vec k$ are constrained to within $1/v_F$.  Thus when we integrate over phase space, either in a scattering process or a loop integral, we do in fact obtain a result suppressed as $g^4 / v_F^3$, which vanishes in the large $v_F$ limit where we fix $g^2 / v_F$.

While a more detailed analysis including higher order contributions is needed, this result suggests that in the limit of large $v_F$ with $g^2 / v_F$ fixed we may consistently neglect effects from Landau damping and from higher scalar correlation functions generated at the quantum level.

\addcontentsline{toc}{section}{References}

\bibliography{shamit}

\begin{thebibliography}{28}%
\makeatletter
\providecommand \@ifxundefined [1]{%
 \@ifx{#1\undefined}
}%
\providecommand \@ifnum [1]{%
 \ifnum #1\expandafter \@firstoftwo
 \else \expandafter \@secondoftwo
 \fi
}%
\providecommand \@ifx [1]{%
 \ifx #1\expandafter \@firstoftwo
 \else \expandafter \@secondoftwo
 \fi
}%
\providecommand \natexlab [1]{#1}%
\providecommand \enquote  [1]{``#1''}%
\providecommand \bibnamefont  [1]{#1}%
\providecommand \bibfnamefont [1]{#1}%
\providecommand \citenamefont [1]{#1}%
\providecommand \href@noop [0]{\@secondoftwo}%
\providecommand \href [0]{\begingroup \@sanitize@url \@href}%
\providecommand \@href[1]{\@@startlink{#1}\@@href}%
\providecommand \@@href[1]{\endgroup#1\@@endlink}%
\providecommand \@sanitize@url [0]{\catcode `\\12\catcode `\$12\catcode
  `\&12\catcode `\#12\catcode `\^12\catcode `\_12\catcode `\%12\relax}%
\providecommand \@@startlink[1]{}%
\providecommand \@@endlink[0]{}%
\providecommand \url  [0]{\begingroup\@sanitize@url \@url }%
\providecommand \@url [1]{\endgroup\@href {#1}{\urlprefix }}%
\providecommand \urlprefix  [0]{URL }%
\providecommand \Eprint [0]{\href }%
\providecommand \doibase [0]{http://dx.doi.org/}%
\providecommand \selectlanguage [0]{\@gobble}%
\providecommand \bibinfo  [0]{\@secondoftwo}%
\providecommand \bibfield  [0]{\@secondoftwo}%
\providecommand \translation [1]{[#1]}%
\providecommand \BibitemOpen [0]{}%
\providecommand \bibitemStop [0]{}%
\providecommand \bibitemNoStop [0]{.\EOS\space}%
\providecommand \EOS [0]{\spacefactor3000\relax}%
\providecommand \BibitemShut  [1]{\csname bibitem#1\endcsname}%
\let\auto@bib@innerbib\@empty
%</preamble>
\bibitem [{\citenamefont {Sachdev}\ \emph {et~al.}(1995)\citenamefont
  {Sachdev}, \citenamefont {Chubukov},\ and\ \citenamefont {Sokol}}]{SCS}%
  \BibitemOpen
  \bibfield  {author} {\bibinfo {author} {\bibfnamefont {S.}~\bibnamefont
  {Sachdev}}, \bibinfo {author} {\bibfnamefont {A.~V.}\ \bibnamefont
  {Chubukov}}, \ and\ \bibinfo {author} {\bibfnamefont {A.}~\bibnamefont
  {Sokol}},\ }\href@noop {} {\bibfield  {journal} {\bibinfo  {journal}
  {Physical Review B}\ }\textbf {\bibinfo {volume} {51}},\ \bibinfo {pages}
  {14874} (\bibinfo {year} {1995})}\BibitemShut {NoStop}%
\bibitem [{\citenamefont {Sachdev}\ and\ \citenamefont
  {Georges}(1995)}]{GeorgesSachdev}%
  \BibitemOpen
  \bibfield  {author} {\bibinfo {author} {\bibfnamefont {S.}~\bibnamefont
  {Sachdev}}\ and\ \bibinfo {author} {\bibfnamefont {A.}~\bibnamefont
  {Georges}},\ }\href@noop {} {\bibfield  {journal} {\bibinfo  {journal}
  {Physical Review B}\ }\textbf {\bibinfo {volume} {52}},\ \bibinfo {pages}
  {9520} (\bibinfo {year} {1995})}\BibitemShut {NoStop}%
\bibitem [{\citenamefont {Mahajan}\ \emph {et~al.}(2013)\citenamefont
  {Mahajan}, \citenamefont {Ramirez}, \citenamefont {Kachru},\ and\
  \citenamefont {Raghu}}]{Mahajan2013}%
  \BibitemOpen
  \bibfield  {author} {\bibinfo {author} {\bibfnamefont {R.}~\bibnamefont
  {Mahajan}}, \bibinfo {author} {\bibfnamefont {D.~M.}\ \bibnamefont
  {Ramirez}}, \bibinfo {author} {\bibfnamefont {S.}~\bibnamefont {Kachru}}, \
  and\ \bibinfo {author} {\bibfnamefont {S.}~\bibnamefont {Raghu}},\ }\href
  {\doibase 10.1103/PhysRevB.88.115116} {\bibfield  {journal} {\bibinfo
  {journal} {Phys. Rev. B}\ }\textbf {\bibinfo {volume} {88}},\ \bibinfo
  {pages} {115116} (\bibinfo {year} {2013})}\BibitemShut {NoStop}%
\bibitem [{\citenamefont {Fitzpatrick}\ \emph {et~al.}(2013)\citenamefont
  {Fitzpatrick}, \citenamefont {Kachru}, \citenamefont {Kaplan},\ and\
  \citenamefont {Raghu}}]{Fitzpatrickone}%
  \BibitemOpen
  \bibfield  {author} {\bibinfo {author} {\bibfnamefont {A.~L.}\ \bibnamefont
  {Fitzpatrick}}, \bibinfo {author} {\bibfnamefont {S.}~\bibnamefont {Kachru}},
  \bibinfo {author} {\bibfnamefont {J.}~\bibnamefont {Kaplan}}, \ and\ \bibinfo
  {author} {\bibfnamefont {S.}~\bibnamefont {Raghu}},\ }\href {\doibase
  10.1103/PhysRevB.88.125116} {\bibfield  {journal} {\bibinfo  {journal} {Phys.
  Rev. B}\ }\textbf {\bibinfo {volume} {88}},\ \bibinfo {pages} {125116}
  (\bibinfo {year} {2013})}\BibitemShut {NoStop}%
\bibitem [{\citenamefont {{Fitzpatrick}}\ \emph {et~al.}(2013)\citenamefont
  {{Fitzpatrick}}, \citenamefont {{Kachru}}, \citenamefont {{Kaplan}},\ and\
  \citenamefont {{Raghu}}}]{FKKRtwo}%
  \BibitemOpen
  \bibfield  {author} {\bibinfo {author} {\bibfnamefont {A.~L.}\ \bibnamefont
  {{Fitzpatrick}}}, \bibinfo {author} {\bibfnamefont {S.}~\bibnamefont
  {{Kachru}}}, \bibinfo {author} {\bibfnamefont {J.}~\bibnamefont {{Kaplan}}},
  \ and\ \bibinfo {author} {\bibfnamefont {S.}~\bibnamefont {{Raghu}}},\
  }\href@noop {} {\bibfield  {journal} {\bibinfo  {journal} {ArXiv e-prints}\ }
  (\bibinfo {year} {2013})},\ \Eprint {http://arxiv.org/abs/1312.3321}
  {arXiv:1312.3321 [cond-mat.str-el]} \BibitemShut {NoStop}%
\bibitem [{\citenamefont {Allais}\ and\ \citenamefont
  {Sachdev}(2014)}]{Allais}%
  \BibitemOpen
  \bibfield  {author} {\bibinfo {author} {\bibfnamefont {A.}~\bibnamefont
  {Allais}}\ and\ \bibinfo {author} {\bibfnamefont {S.}~\bibnamefont
  {Sachdev}},\ }\href@noop {} {\bibfield  {journal} {\bibinfo  {journal}
  {Physical Review B}\ }\textbf {\bibinfo {volume} {90}},\ \bibinfo {pages}
  {035131} (\bibinfo {year} {2014})}\BibitemShut {NoStop}%
\bibitem [{\citenamefont {Son}(1999)}]{Son1999}%
  \BibitemOpen
  \bibfield  {author} {\bibinfo {author} {\bibfnamefont {D.~T.}\ \bibnamefont
  {Son}},\ }\href {\doibase 10.1103/PhysRevD.59.094019} {\bibfield  {journal}
  {\bibinfo  {journal} {Phys. Rev. D}\ }\textbf {\bibinfo {volume} {59}},\
  \bibinfo {pages} {094019} (\bibinfo {year} {1999})}\BibitemShut {NoStop}%
\bibitem [{\citenamefont {Shuster}\ and\ \citenamefont {Son}(2000)}]{Shuster}%
  \BibitemOpen
  \bibfield  {author} {\bibinfo {author} {\bibfnamefont {E.}~\bibnamefont
  {Shuster}}\ and\ \bibinfo {author} {\bibfnamefont {D.}~\bibnamefont {Son}},\
  }\href {\doibase 10.1016/S0550-3213(99)00615-X} {\bibfield  {journal}
  {\bibinfo  {journal} {Nucl.Phys.}\ }\textbf {\bibinfo {volume} {B573}},\
  \bibinfo {pages} {434} (\bibinfo {year} {2000})},\ \Eprint
  {http://arxiv.org/abs/hep-ph/9905448} {arXiv:hep-ph/9905448 [hep-ph]}
  \BibitemShut {NoStop}%
%%CITATION = HEP-PH/9905448;%%
\bibitem [{\citenamefont {Chubukov}\ and\ \citenamefont
  {Schmalian}(2005)}]{Chubukov2005}%
  \BibitemOpen
  \bibfield  {author} {\bibinfo {author} {\bibfnamefont {A.~V.}\ \bibnamefont
  {Chubukov}}\ and\ \bibinfo {author} {\bibfnamefont {J.}~\bibnamefont
  {Schmalian}},\ }\href@noop {} {\bibfield  {journal} {\bibinfo  {journal}
  {Physical Review B}\ }\textbf {\bibinfo {volume} {72}},\ \bibinfo {pages}
  {174520} (\bibinfo {year} {2005})}\BibitemShut {NoStop}%
\bibitem [{\citenamefont {She}\ and\ \citenamefont
  {Zaanen}(2009)}]{Zaanen2008}%
  \BibitemOpen
  \bibfield  {author} {\bibinfo {author} {\bibfnamefont {J.-H.}\ \bibnamefont
  {She}}\ and\ \bibinfo {author} {\bibfnamefont {J.}~\bibnamefont {Zaanen}},\
  }\href {\doibase 10.1103/PhysRevB.80.184518} {\bibfield  {journal} {\bibinfo
  {journal} {Phys. Rev. B}\ }\textbf {\bibinfo {volume} {80}},\ \bibinfo
  {pages} {184518} (\bibinfo {year} {2009})}\BibitemShut {NoStop}%
\bibitem [{\citenamefont {Moon}\ and\ \citenamefont
  {Chubukov}(2010)}]{Moon2010}%
  \BibitemOpen
  \bibfield  {author} {\bibinfo {author} {\bibfnamefont {E.-G.}\ \bibnamefont
  {Moon}}\ and\ \bibinfo {author} {\bibfnamefont {A.}~\bibnamefont
  {Chubukov}},\ }\href@noop {} {\bibfield  {journal} {\bibinfo  {journal}
  {Journal of Low Temperature Physics}\ }\textbf {\bibinfo {volume} {161}},\
  \bibinfo {pages} {263} (\bibinfo {year} {2010})}\BibitemShut {NoStop}%
\bibitem [{\citenamefont {Metlitski}\ \emph {et~al.}(2014)\citenamefont
  {Metlitski}, \citenamefont {Mross}, \citenamefont {Sachdev},\ and\
  \citenamefont {Senthil}}]{Metlitski}%
  \BibitemOpen
  \bibfield  {author} {\bibinfo {author} {\bibfnamefont {M.~A.}\ \bibnamefont
  {Metlitski}}, \bibinfo {author} {\bibfnamefont {D.~F.}\ \bibnamefont
  {Mross}}, \bibinfo {author} {\bibfnamefont {S.}~\bibnamefont {Sachdev}}, \
  and\ \bibinfo {author} {\bibfnamefont {T.}~\bibnamefont {Senthil}},\
  }\href@noop {} {\bibfield  {journal} {\bibinfo  {journal} {arXiv preprint
  arXiv:1403.3694}\ } (\bibinfo {year} {2014})}\BibitemShut {NoStop}%
\bibitem [{\citenamefont {{Maier}}\ and\ \citenamefont
  {{Scalapino}}(2014)}]{Maier2014}%
  \BibitemOpen
  \bibfield  {author} {\bibinfo {author} {\bibfnamefont {T.~A.}\ \bibnamefont
  {{Maier}}}\ and\ \bibinfo {author} {\bibfnamefont {D.~J.}\ \bibnamefont
  {{Scalapino}}},\ }\href@noop {} {\bibfield  {journal} {\bibinfo  {journal}
  {ArXiv e-prints}\ } (\bibinfo {year} {2014})},\ \Eprint
  {http://arxiv.org/abs/1405.5238} {arXiv:1405.5238 [cond-mat.supr-con]}
  \BibitemShut {NoStop}%
\bibitem [{\citenamefont {Nayak}\ and\ \citenamefont
  {Wilczek}(1994)}]{Nayak1994a}%
  \BibitemOpen
  \bibfield  {author} {\bibinfo {author} {\bibfnamefont {C.}~\bibnamefont
  {Nayak}}\ and\ \bibinfo {author} {\bibfnamefont {F.}~\bibnamefont
  {Wilczek}},\ }\href@noop {} {\bibfield  {journal} {\bibinfo  {journal}
  {Nuclear Physics B}\ }\textbf {\bibinfo {volume} {430}},\ \bibinfo {pages}
  {534} (\bibinfo {year} {1994})}\BibitemShut {NoStop}%
\bibitem [{\citenamefont {Sch\"afer}\ and\ \citenamefont
  {Wilczek}(1999)}]{Schafer1999}%
  \BibitemOpen
  \bibfield  {author} {\bibinfo {author} {\bibfnamefont {T.}~\bibnamefont
  {Sch\"afer}}\ and\ \bibinfo {author} {\bibfnamefont {F.}~\bibnamefont
  {Wilczek}},\ }\href {\doibase 10.1103/PhysRevD.60.114033} {\bibfield
  {journal} {\bibinfo  {journal} {Phys. Rev. D}\ }\textbf {\bibinfo {volume}
  {60}},\ \bibinfo {pages} {114033} (\bibinfo {year} {1999})}\BibitemShut
  {NoStop}%
\bibitem [{\citenamefont {Wang}\ and\ \citenamefont
  {Chubukov}(2013)}]{Wang2013}%
  \BibitemOpen
  \bibfield  {author} {\bibinfo {author} {\bibfnamefont {Y.}~\bibnamefont
  {Wang}}\ and\ \bibinfo {author} {\bibfnamefont {A.~V.}\ \bibnamefont
  {Chubukov}},\ }\href {\doibase 10.1103/PhysRevLett.110.127001} {\bibfield
  {journal} {\bibinfo  {journal} {Phys. Rev. Lett.}\ }\textbf {\bibinfo
  {volume} {110}},\ \bibinfo {pages} {127001} (\bibinfo {year}
  {2013})}\BibitemShut {NoStop}%
\bibitem [{\citenamefont {Dalidovich}\ and\ \citenamefont
  {Lee}(2013)}]{Lee2013}%
  \BibitemOpen
  \bibfield  {author} {\bibinfo {author} {\bibfnamefont {D.}~\bibnamefont
  {Dalidovich}}\ and\ \bibinfo {author} {\bibfnamefont {S.-S.}\ \bibnamefont
  {Lee}},\ }\href {\doibase 10.1103/PhysRevB.88.245106} {\bibfield  {journal}
  {\bibinfo  {journal} {Phys. Rev. B}\ }\textbf {\bibinfo {volume} {88}},\
  \bibinfo {pages} {245106} (\bibinfo {year} {2013})}\BibitemShut {NoStop}%
\bibitem [{\citenamefont {Lee}(2009)}]{Lee2009}%
  \BibitemOpen
  \bibfield  {author} {\bibinfo {author} {\bibfnamefont {S.-S.}\ \bibnamefont
  {Lee}},\ }\href {\doibase 10.1103/PhysRevB.80.165102} {\bibfield  {journal}
  {\bibinfo  {journal} {Phys. Rev. B}\ }\textbf {\bibinfo {volume} {80}},\
  \bibinfo {pages} {165102} (\bibinfo {year} {2009})}\BibitemShut {NoStop}%
\bibitem [{\citenamefont {Fitzpatrick}\ \emph {et~al.}(2014)\citenamefont
  {Fitzpatrick}, \citenamefont {Torroba},\ and\ \citenamefont {Wang}}]{WTF}%
  \BibitemOpen
  \bibfield  {author} {\bibinfo {author} {\bibfnamefont {A.~L.}\ \bibnamefont
  {Fitzpatrick}}, \bibinfo {author} {\bibfnamefont {G.}~\bibnamefont
  {Torroba}}, \ and\ \bibinfo {author} {\bibfnamefont {H.}~\bibnamefont
  {Wang}},\ }\href@noop {} {\  (\bibinfo {year} {2014})},\ \Eprint
  {http://arxiv.org/abs/1410.6811} {arXiv:1410.6811 [cond-mat.str-el]}
  \BibitemShut {NoStop}%
%%CITATION = ARXIV:1410.6811;%%
\bibitem [{\citenamefont {Torroba}\ and\ \citenamefont {Wang}(2014)}]{Gonzalo}%
  \BibitemOpen
  \bibfield  {author} {\bibinfo {author} {\bibfnamefont {G.}~\bibnamefont
  {Torroba}}\ and\ \bibinfo {author} {\bibfnamefont {H.}~\bibnamefont {Wang}},\
  }\href {\doibase 10.1103/PhysRevB.90.165144} {\bibfield  {journal} {\bibinfo
  {journal} {Phys.Rev.}\ }\textbf {\bibinfo {volume} {B90}},\ \bibinfo {pages}
  {165144} (\bibinfo {year} {2014})},\ \Eprint {http://arxiv.org/abs/1406.3029}
  {arXiv:1406.3029 [cond-mat.str-el]} \BibitemShut {NoStop}%
%%CITATION = ARXIV:1406.3029;%%
\bibitem [{\citenamefont {{Yamamoto}}\ and\ \citenamefont
  {{Si}}(2010)}]{Yamamoto2010}%
  \BibitemOpen
  \bibfield  {author} {\bibinfo {author} {\bibfnamefont {S.~J.}\ \bibnamefont
  {{Yamamoto}}}\ and\ \bibinfo {author} {\bibfnamefont {Q.}~\bibnamefont
  {{Si}}},\ }\href {\doibase 10.1103/PhysRevB.81.205106} {\bibfield  {journal}
  {\bibinfo  {journal} {Phys.Rev.}\ }\textbf {\bibinfo {volume} {B81}},\
  \bibinfo {pages} {205106} (\bibinfo {year} {2010})},\ \Eprint
  {http://arxiv.org/abs/0906.0014} {arXiv:0906.0014 [cond-mat.str-el]}
  \BibitemShut {NoStop}%
\bibitem [{\citenamefont {Shankar}(1994)}]{Shankar}%
  \BibitemOpen
  \bibfield  {author} {\bibinfo {author} {\bibfnamefont {R.}~\bibnamefont
  {Shankar}},\ }\href {\doibase 10.1103/RevModPhys.66.129} {\bibfield
  {journal} {\bibinfo  {journal} {Rev.Mod.Phys.}\ }\textbf {\bibinfo {volume}
  {66}},\ \bibinfo {pages} {129} (\bibinfo {year} {1994})}\BibitemShut
  {NoStop}%
%%CITATION = RMPHA,66,129;%%
\bibitem [{\citenamefont {Polchinski}(1992)}]{Polchinski}%
  \BibitemOpen
  \bibfield  {author} {\bibinfo {author} {\bibfnamefont {J.}~\bibnamefont
  {Polchinski}},\ }\href@noop {} {\  (\bibinfo {year} {1992})},\ \Eprint
  {http://arxiv.org/abs/hep-th/9210046} {arXiv:hep-th/9210046 [hep-th]}
  \BibitemShut {NoStop}%
%%CITATION = HEP-TH/9210046;%%
\bibitem [{Note1()}]{Note1}%
  \BibitemOpen
  \bibinfo {note} {The decimation of high momentum bosons through variations of
  $\Lambda _b$ is also important for canceling log divergences in nonlocal
  operators, thus ensuring the renormalizability of the theory [\protect
  \rev@citealpnum {WTF}].}\BibitemShut {Stop}%
\bibitem [{Note2()}]{Note2}%
  \BibitemOpen
  \bibinfo {note} {Ref. \protect \rev@citealpnum {Gonzalo} confirmed that the
  $\beta $ functions and anomalous dimensions in the theory are dominated by
  regions of the relevant loop integrals when the boson $q_0$ is ${\protect
  \cal O}(q)$, rather than ${\protect \cal O}(v q)$, so holding $x$ fixed and
  ${\protect \cal O}(1)$ is the correct limit to take.}\BibitemShut {Stop}%
\bibitem [{tre(pear)}]{treepee}%
  \BibitemOpen
  \href@noop {} {\  (\bibinfo {year} {To appear})}\BibitemShut {NoStop}%
\bibitem [{\citenamefont {Peierls}(1930)}]{Peierls1930}%
  \BibitemOpen
  \bibfield  {author} {\bibinfo {author} {\bibfnamefont {R.}~\bibnamefont
  {Peierls}},\ }\href@noop {} {\bibfield  {journal} {\bibinfo  {journal} {Ann.
  Phys. Leipzig}\ }\textbf {\bibinfo {volume} {4}},\ \bibinfo {pages} {121}
  (\bibinfo {year} {1930})}\BibitemShut {NoStop}%
\bibitem [{\citenamefont {Deryagin}\ \emph {et~al.}(1992)\citenamefont
  {Deryagin}, \citenamefont {Grigoriev},\ and\ \citenamefont {Rubakov}}]{DGR}%
  \BibitemOpen
  \bibfield  {author} {\bibinfo {author} {\bibfnamefont {D.}~\bibnamefont
  {Deryagin}}, \bibinfo {author} {\bibfnamefont {D.~Y.}\ \bibnamefont
  {Grigoriev}}, \ and\ \bibinfo {author} {\bibfnamefont {V.}~\bibnamefont
  {Rubakov}},\ }\href {\doibase 10.1142/S0217751X92000302} {\bibfield
  {journal} {\bibinfo  {journal} {Int.J.Mod.Phys.}\ }\textbf {\bibinfo {volume}
  {A7}},\ \bibinfo {pages} {659} (\bibinfo {year} {1992})}\BibitemShut
  {NoStop}%
%%CITATION = IMPAE,A7,659;%%
\end{thebibliography}%

\end{document}